\documentclass[10pt, conference, letterpaper]{IEEEtran}

\usepackage{url}
\usepackage[pdftex]{graphicx}
\DeclareGraphicsExtensions{.pdf, .png}
\usepackage{booktabs}
\usepackage{tabularx}
\usepackage{float}

\usepackage{placeins}
\usepackage{multirow}
\usepackage{color}

\usepackage{units}

\usepackage{cite}
\usepackage{amsfonts}
\usepackage{amssymb}
\usepackage[cmex10]{amsmath}
\usepackage{array}
\usepackage{mdwmath}
\usepackage{mdwtab}
\ifCLASSOPTIONcompsoc
\usepackage[caption=false, font=normalsize, labelfont=sf, textfont=sf] {subfig}
\else
\usepackage[caption=false, font=footnotesize]{subfig}
\fi
\begin{document}
%\title{60~GHz Outdoor Urban Measurement Study of Antenna Beamsteering Requirements for Millimeter-Wave Cellular Networks}
\title{60~GHz Outdoor Urban Measurement Study\\of the Feasibility of Multi-Gbps
\\mm-Wave Cellular Networks}
\author{\IEEEauthorblockN{Ljiljana Simi\'{c}, Nikos Perpinias and Marina Petrova\\
Institute for Networked Systems, RWTH Aachen University\\
Kackertstrasse 9, D-52072 Aachen, Germany\\%}}
E-mail: \{lsi, npe, mpe\}@inets.rwth-aachen.de}}
\maketitle

\begin{abstract}
\boldmath
Future 5G cellular networks are expected to exploit the abundant spectrum resources of the millimeter-wave (mm-wave) bands to satisfy demand for multi-Gbps mobile links anticipated by exponential data traffic growth. However, given the directional nature of mm-wave links, the feasibility of mm-wave mobile \emph{networks} is critically dependent on efficient antenna beamsteering and a rich inventory of strong LOS (line-of-sight) and NLOS (non line-of-sight) paths from effective reflectors in the urban environment. In this paper we report results from detailed angular measurements of $60~\text{GHz}$ links at an example outdoor pico-cellular site in a mixed-use urban environment typical of European cities. Our work is the first to systematically analyze the beamsteering requirements of future mm-wave cellular networks based on real measurements. Our results reveal that the urban environment provides substantial opportunities for multi-Gbps mm-wave connectivity, but that the availability of strong LOS/NLOS links is highly location and orientation-specific. Our results also show that high speed mm-wave links are very sensitive to beam misalignment. This has important implications for practical mm-wave cellular network design: (i) high-precision beamsteering is required to maintain stable data rates even for quasi-stationary users; and (ii) providing seamless high speed service in mobility scenarios will be extremely challenging. Our results thus cast doubt on whether outdoor mm-wave cellular deployments will be feasible in practice, given the high network control overhead of meeting such stringent beamsteering requirements. 
%\begin{keywords}
%Millimeter-wave, 60~GHz, measurements, urban outdoor cellular network, IEEE 802.11ad
%\end{keywords}
\end{abstract}
%\IEEEpeerreviewmaketitle

%------------------------------------------------------------
%- c.f. TED: looking at estimating range and rough coverage possibility, we instead focus on CLOSER/STRONGER links in hotspot/micropico-cellular scenario (max radius of our BS is under  30 m) \& see if even for this less (pathloss-challenged)challenging case, we can/appears feasibly we can provide seamless Gbps coverage \& STABLE service to MOBILE users.
%------------------------------------------------------------
%===========================================================================================================================
\section{Introduction}

%PLUS NOW HAVE TECHNOLOGY TO FABRICATE)
The plentiful spectrum resources available at the millimeter-wave (mm-wave) bands are widely expected to be a key means of addressing the challenges of exponential mobile data growth -- and the resulting ``spectrum crunch'' in the current licensed microwave bands --  in future 5G cellular networks~\cite{Andrews2014_whatWill5GBe,Rappaport2015_itWillWork}. However, outdoor propagation at mm-wave frequencies is challenging. Firstly, it relies on the existence of LOS (line-of-sight) or strong reflected NLOS (non line-of-sight) transmission paths~\cite{Molisch2014_PropChannelModelsNextG}. This is in contrast to traditional cellular networks where diffraction is a major propagation mechanism facilitating coverage in NLOS %urban
environments. Namely, link opportunities in mm-wave networks are far more dependent on the idiosyncratic urban layout at a cell site than in existing cellular deployments. Secondly, overcoming the inherently high path loss at mm-wave necessitates the use of directional antennas. Accordingly, the added complexity of performing beamsteering to establish and maintain strong LOS/NLOS links for mobile users is likely to impose a significant control and signalling overhead in mm-wave cellular networks. 

Detailed \emph{directional} characterization of mm-wave connectivity is thus crucial for gaining insight into the beamsteering requirements and opportunities of outdoor mm-wave networks, but is largely unaddressed in the existing literature~\cite{Molisch2014_PropChannelModelsNextG}. The seminal outdoor measurements conducted by Rappaport \emph{et~al.}~\cite{Rappaport2015_MeasurmentsSurvey} using a wideband sliding correlator channel sounder, at $28~\text{GHz}$~\cite{Samimi2013_28GHz_AoA} and $73~\text{GHz}$~\cite{MacCartney2014_73GHz} in New York City and at $38~\text{GHz}$ and $60~\text{GHz}~$\cite{Rappaport2012_3860GHz} in Austin, have shown that mm-wave links are possible up to a cell radius of $200~\text{m}$. In~\cite{Samimi2013_28GHz_AoA}, the authors present illustrative AoA (angle of arrival) results, in the azimuth plane only, demonstrating that on average three main ``lobes'' exist corresponding to distinct LOS/NLOS paths.  However, the work of Rappaport \emph{et al.} has otherwise largely focused on fundamental mm-wave propagation channel modelling, including time-domain characterization and empirically fitting average LOS/NLOS path loss exponents to provide initial estimates of the coverage achievable with mm-wave links.% rather than systematic sweeps.

By contrast, in this paper we take a system-level, network design perspective to assessing the feasibility of mm-wave cellular networks. We address the key question of: how stringent are the beamsteering requirements in urban mm-wave networks for seamless multi-Gbps mobile data provisioning? We report results from a $60~\text{GHz}$ outdoor urban measurement study in the centre of a typical European city, and present an analysis of the fine-grained angular characteristics of mm-wave links at our example pico-cellular site. Our results reveal that the urban environment considered provides significant multi-Gbps link opportunities, but that high-precision beamsteering is needed to maintain a stable high data rate, even for a quasi-stationary user.  Namely, our results suggest that the beamsteering complexity may be prohibitively high to make mm-wave deployments attractive in practice for outdoor mobile environments.

To the best of our knowledge, our work is the first to systematically study the beamsteering requirements for future mm-wave cellular networks based on real outdoor urban measurements. We are also the first to report detailed measurements for a $60~\text{GHz}$ cellular-like outdoor urban deployment: only peer-to-peer measurements were reported at $60~\text{GHz}$ in~\cite{Rappaport2012_3860GHz}, the Berlin measurements in~\cite{FraunhoferBerlin2014} solely characterized LOS links with omnidirectional antennas, and~\cite{Heather2014_demystifying} considered only conceptual networking aspects. Moreover, existing mm-wave outdoor measurements~\cite{Rappaport2015_MeasurmentsSurvey, FraunhoferBerlin2014} have been in urban environments characteristic of very large cities and CBD-like urban layouts, with wide streets and modern buildings made of relatively homogeneous materials (typically highly reflective, e.g. concrete and glass). We instead consider a mixed-use urban environment with heterogenous building types and materials, typical of most European cities, and quantify the extent of reflected NLOS opportunities with respect to the urban layout and materials. Finally, given the general lack of experimental studies in mm-wave networking, we believe our methodology will also be of interest to the emerging experimental indoor $60~\text{GHz}$ research community~\cite{Widmer2015, Arnold2015_SRIF}.%, where the first experimental works are only recently emerging~\cite{Widmer2015, Arnold2015_SRIF}.

The rest of this paper is organized as follows. Section~\ref{SECmeasurementSetup} describes our measurement setup and methodology. In Section~\ref{SECresults} we present and analyze our measurement results. Section~\ref{SECconclusions} concludes the paper. 

%\clearpage
%%%....................................................................................
%\begin{figure}[!tb]
%\centering
%\subfloat[all RX locations]{\includegraphics[width=0.48\columnwidth]{map_ALL}
%\label{fig:MAPS_all}}
%\hspace{1mm}\subfloat[RX position A]{\includegraphics[width=0.48\columnwidth]{point_A}
%\label{fig:MAPS_A}}
%\\\subfloat[RX position B]{\includegraphics[width=0.48\columnwidth]{point_B}
%\label{fig:MAPS_B}}
%\hspace{1mm}\subfloat[RX position C]{\includegraphics[width=0.48\columnwidth]{point_C}
%\label{fig:MAPS_C}}
%\\\subfloat[RX position D]{\includegraphics[width=0.48\columnwidth]{point_D}
%\label{fig:MAPS_D}}
%\hspace{1mm}\subfloat[RX position E]{\includegraphics[width=0.48\columnwidth]{point_E}
%\label{fig:MAPS_E}}
%\\\subfloat[RX position F]{\includegraphics[width=0.48\columnwidth]{point_F}
%\label{fig:MAPS_F}}
%\hspace{1mm}\subfloat[RX position G]{\includegraphics[width=0.48\columnwidth]{point_G}
%\label{fig:MAPS_G}}
%\caption{\small{Maps illustrating the measurement locations, showing for each receiver location the major azimuth direction of LOS and NLOS links found with respect to surrounding building structures and materials.}}
%\label{fig:MAPS}
%\end{figure}
%%....................................................................................
%%....................................................................................
\begin{figure*}[!tb]
\centering
\subfloat[all RX locations]{\includegraphics[width=0.48\columnwidth]{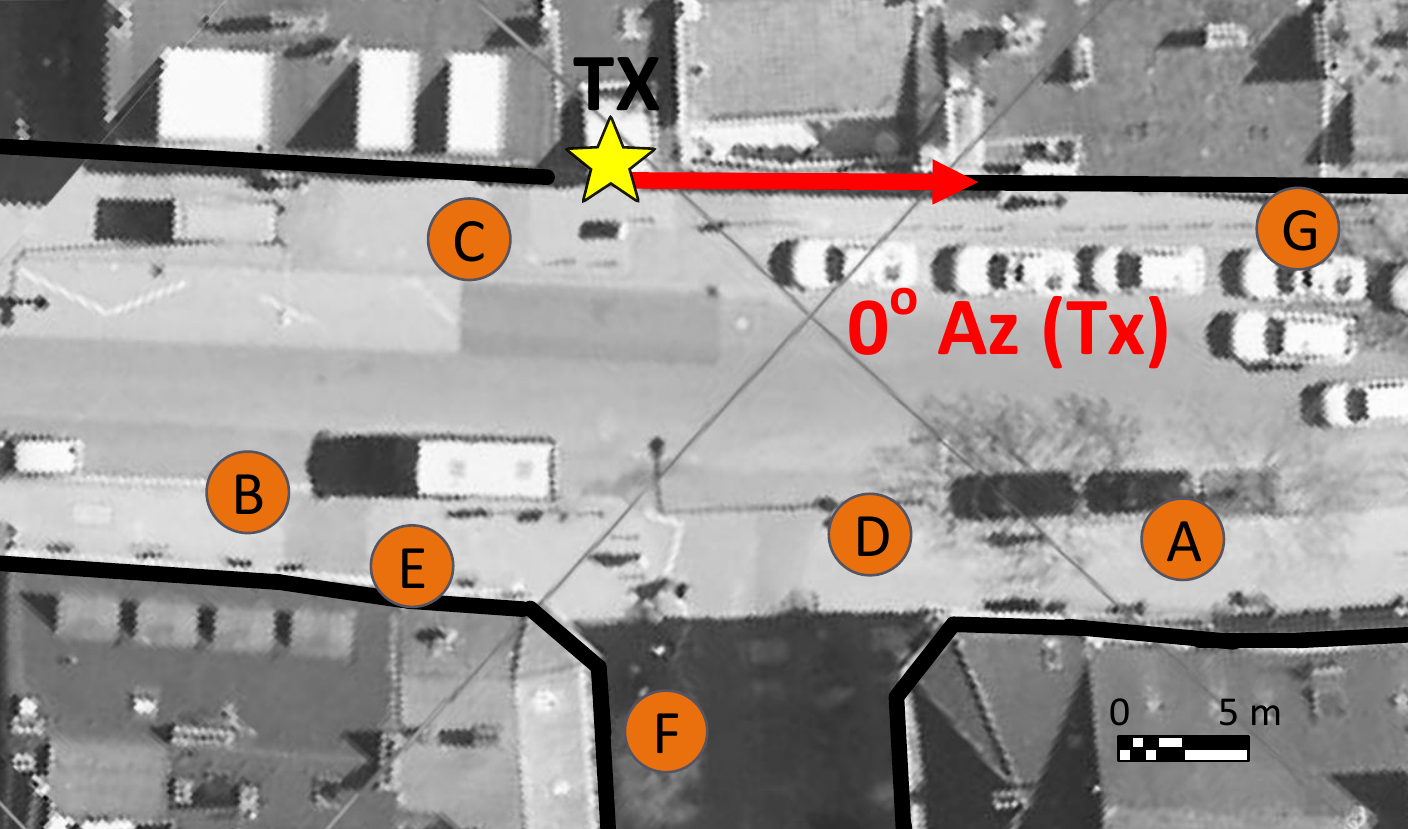}
\label{fig:MAPS_all}}
\hspace{1mm}\subfloat[RX location A]{\includegraphics[width=0.48\columnwidth]{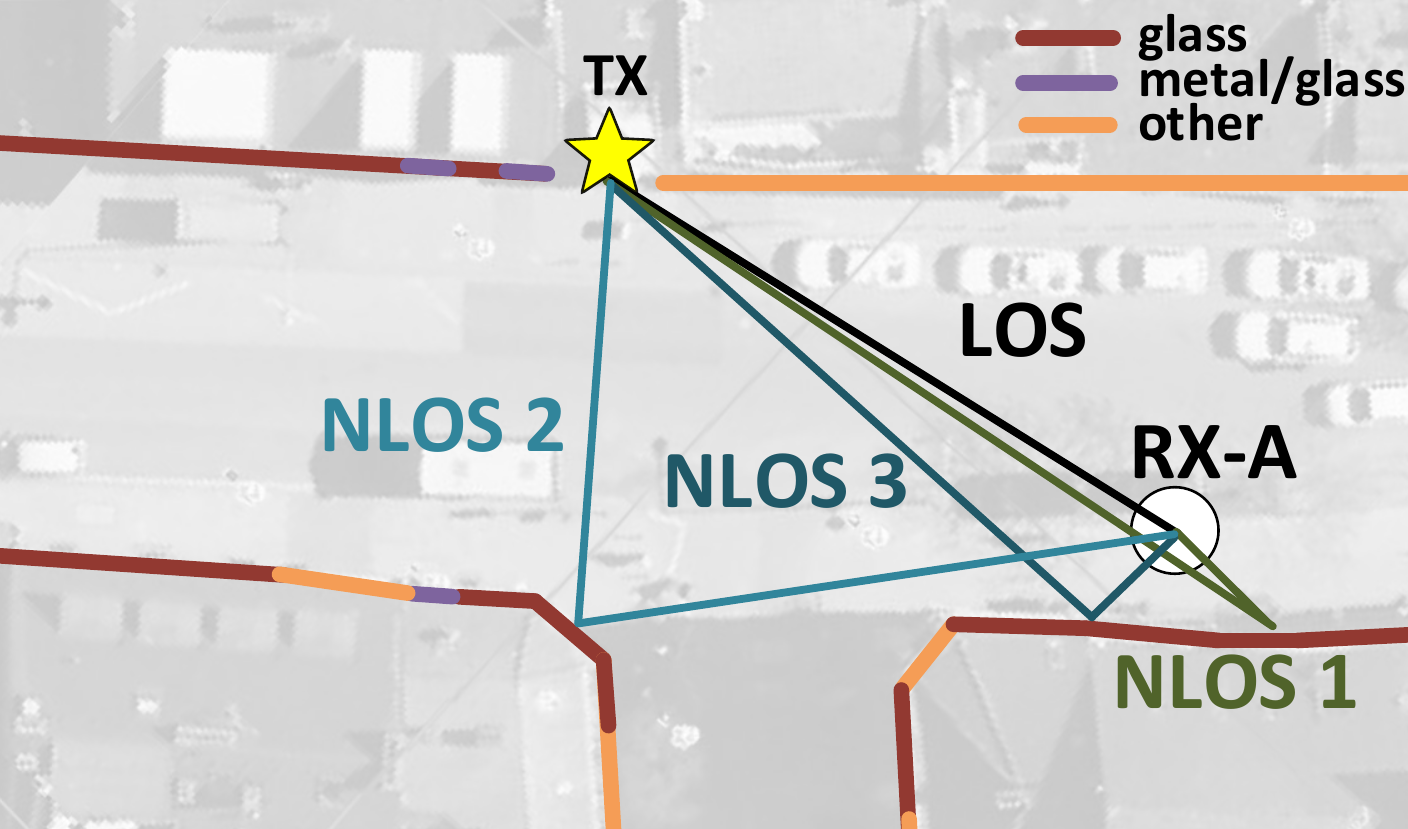}
\label{fig:MAPS_A}}
\hspace{1mm}\subfloat[RX location B]{\includegraphics[width=0.48\columnwidth]{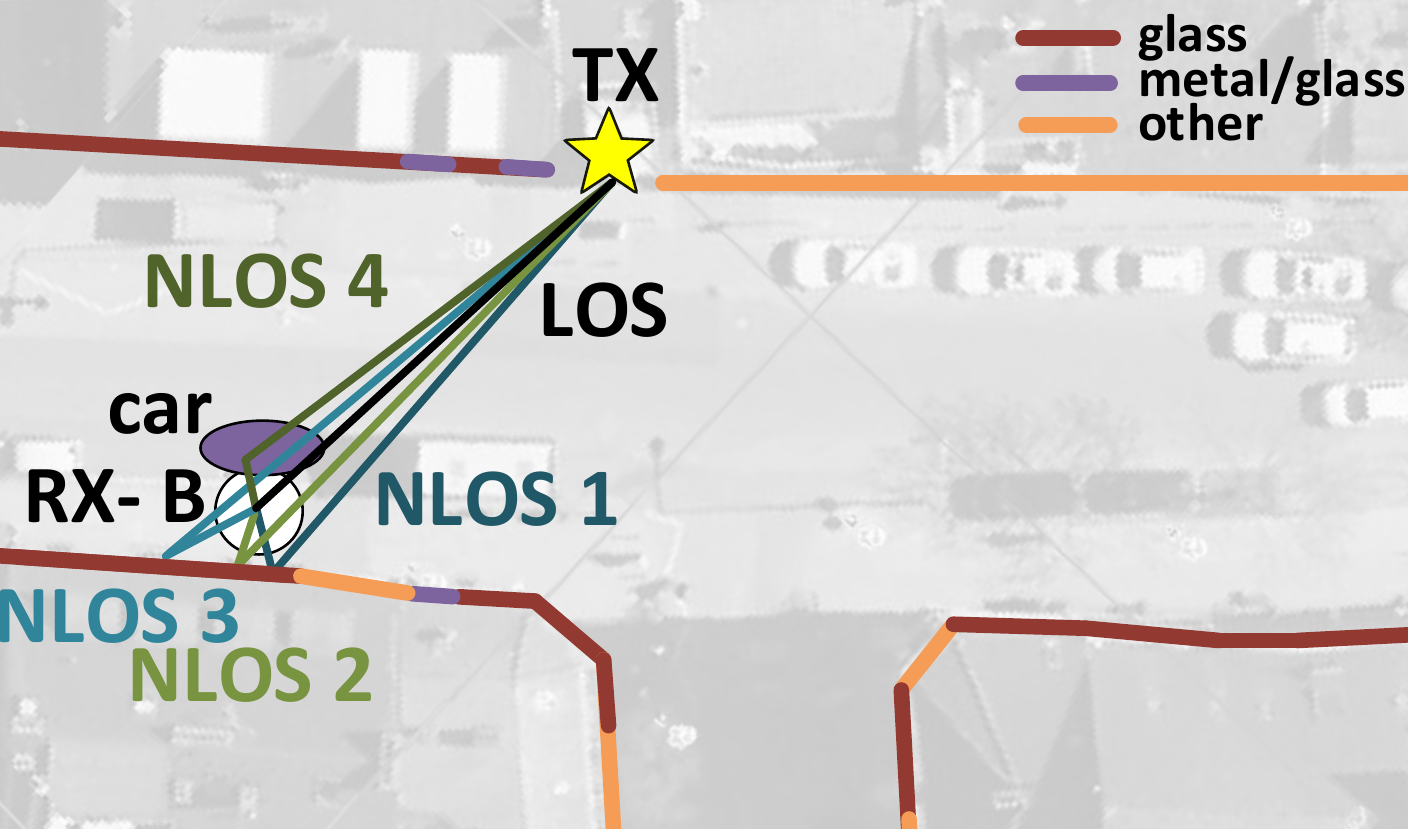}
\label{fig:MAPS_B}}
\hspace{1mm}\subfloat[RX location C]{\includegraphics[width=0.48\columnwidth]{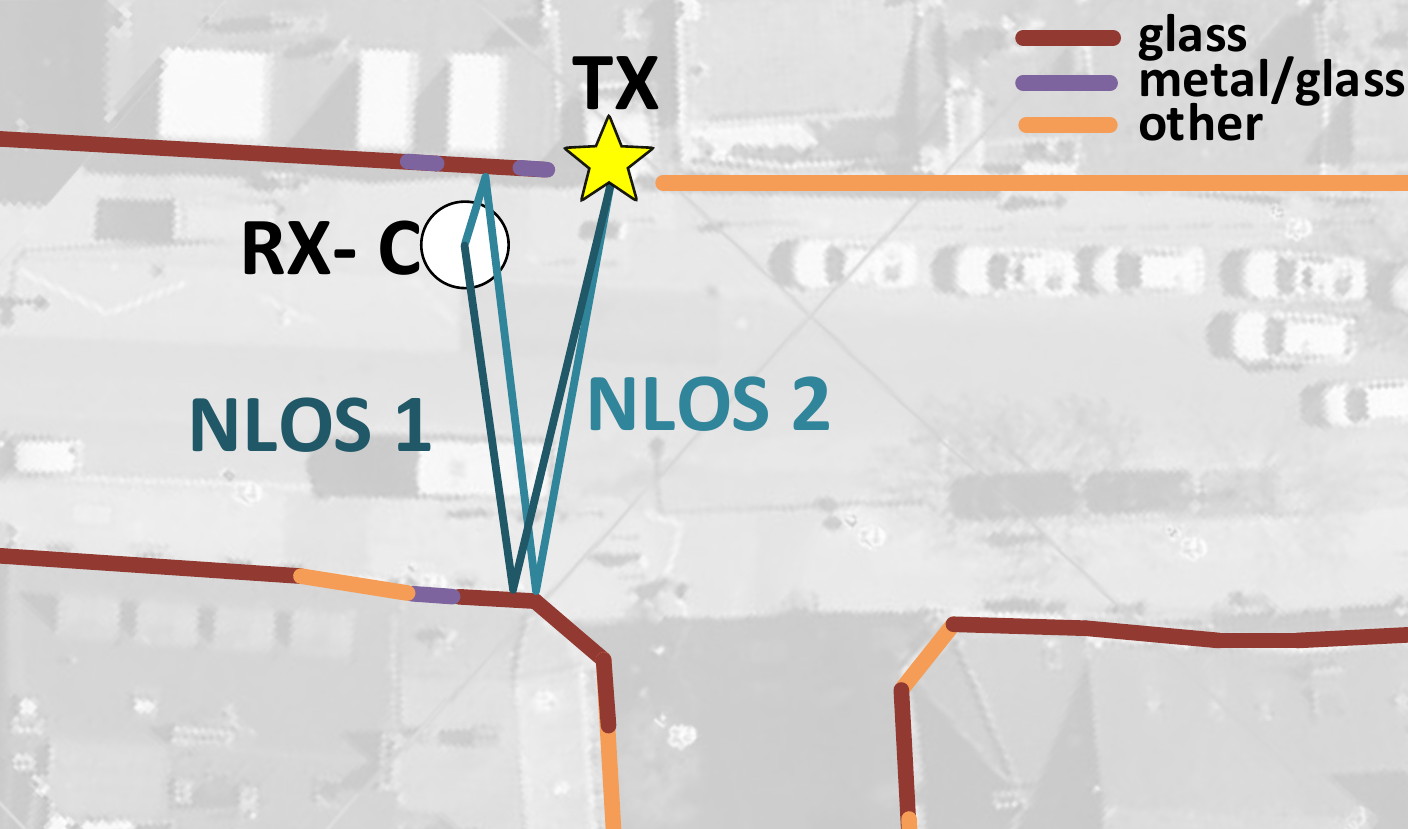}
\label{fig:MAPS_C}}
\\\subfloat[RX location D]{\includegraphics[width=0.48\columnwidth]{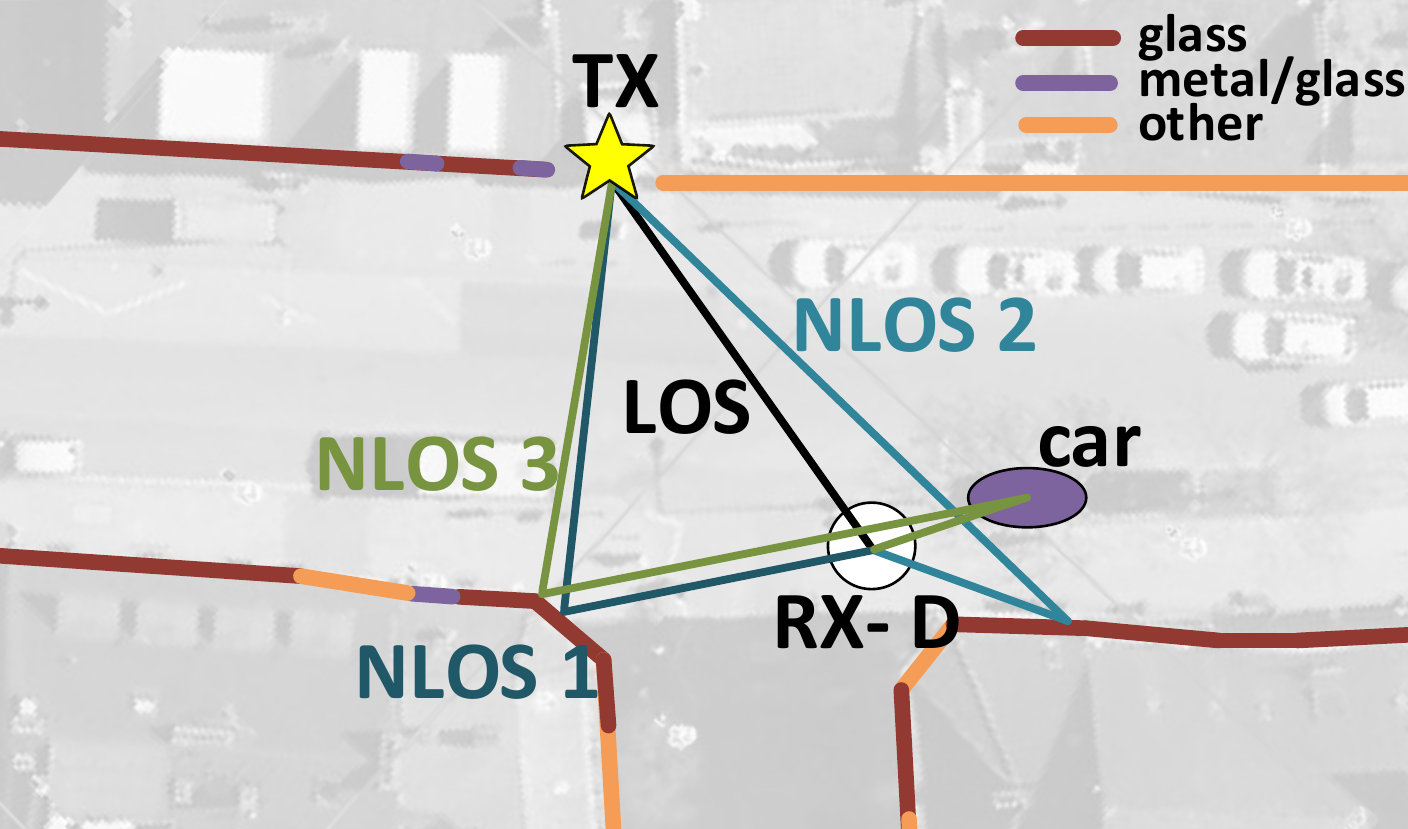}
\label{fig:MAPS_D}}
\hspace{1mm}\subfloat[RX location E]{\includegraphics[width=0.48\columnwidth]{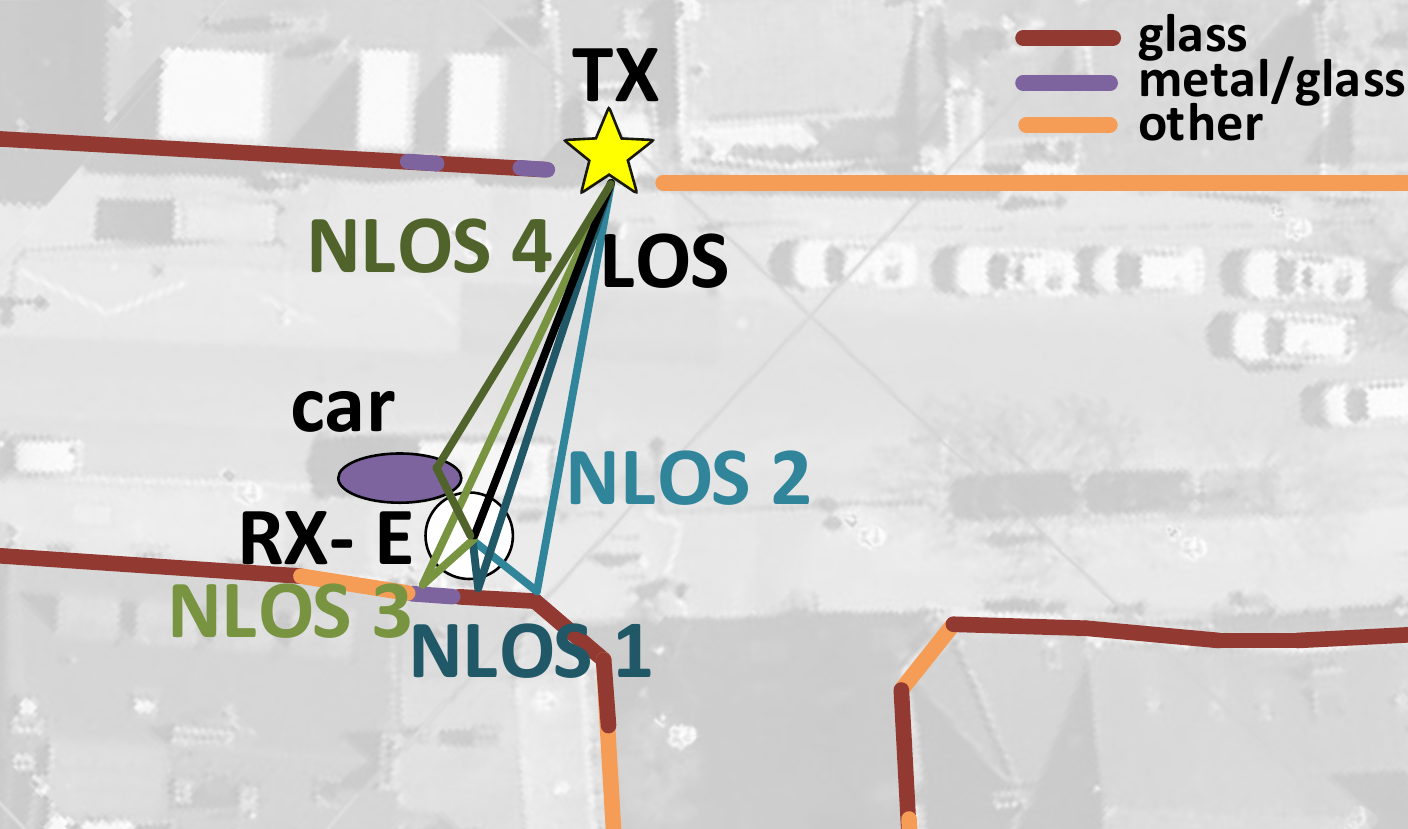}
\label{fig:MAPS_E}}
\hspace{1mm}\subfloat[RX location F]{\includegraphics[width=0.48\columnwidth]{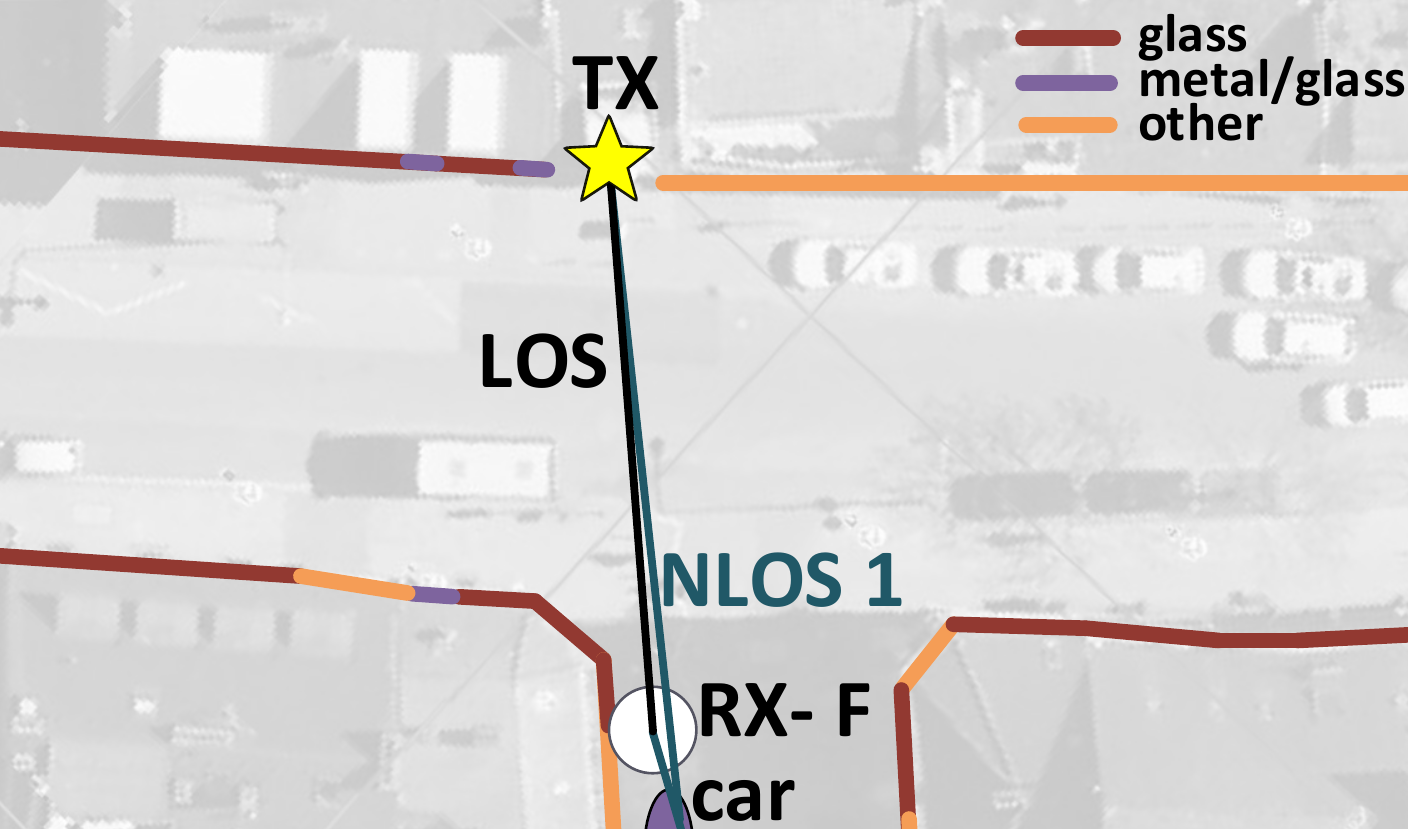}
\label{fig:MAPS_F}}
\hspace{1mm}\hspace{1mm}\subfloat[RX location G]{\includegraphics[width=0.48\columnwidth]{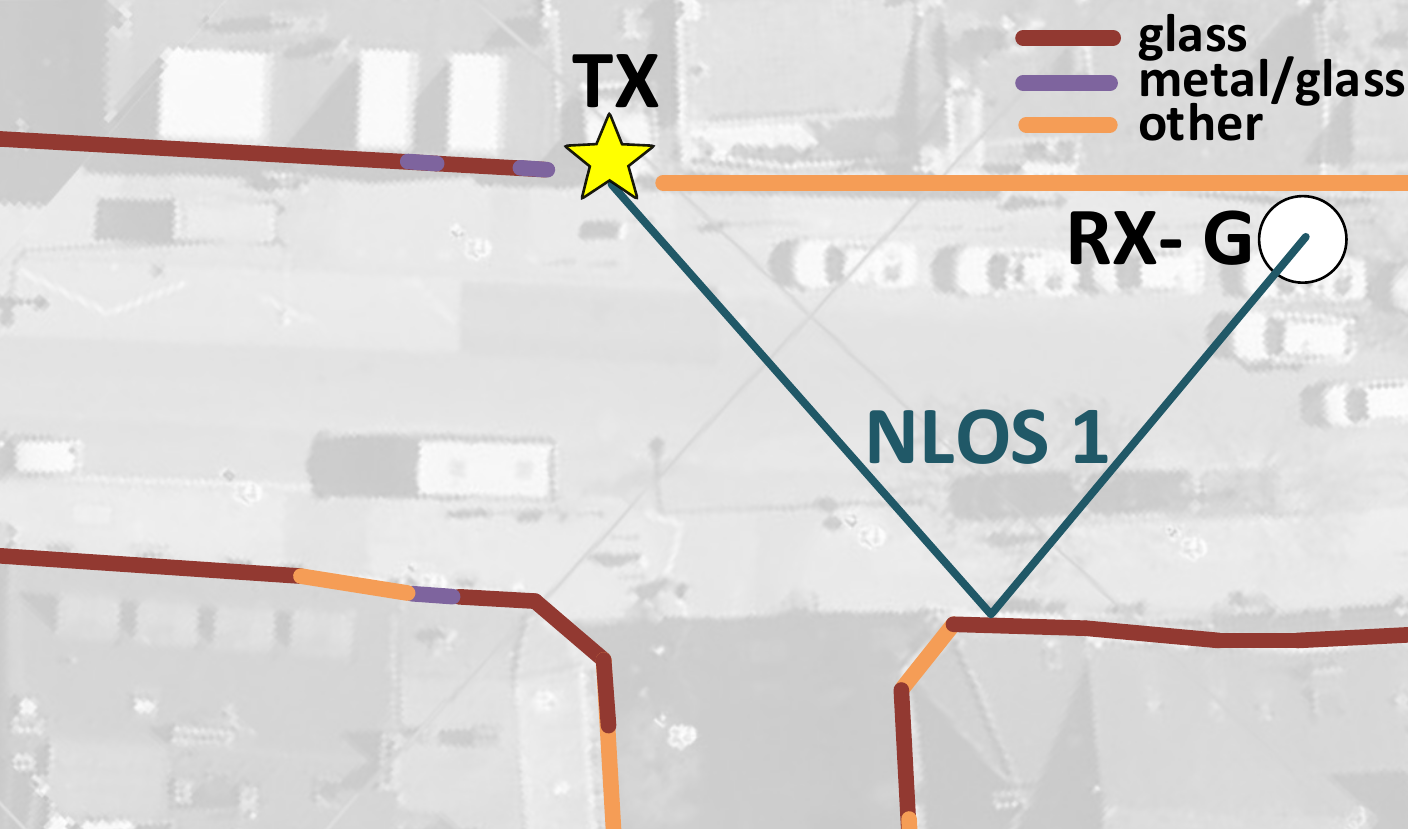}
\label{fig:MAPS_G}}
\caption{\small{Maps illustrating the mixed-use urban measurement study area, showing (a) the locations of the transmitter (TX) and receiver (RX) and (b)-(h) for each RX location the azimuth direction of major LOS/NLOS links found w.r.t. surrounding building layout and materials.}}
\label{fig:MAPS}
\end{figure*}
%....................................................................................
%....................................................................................
\begin{figure*}[!tb]
\centering
\subfloat[TX block diagram]{\includegraphics[scale=0.27]{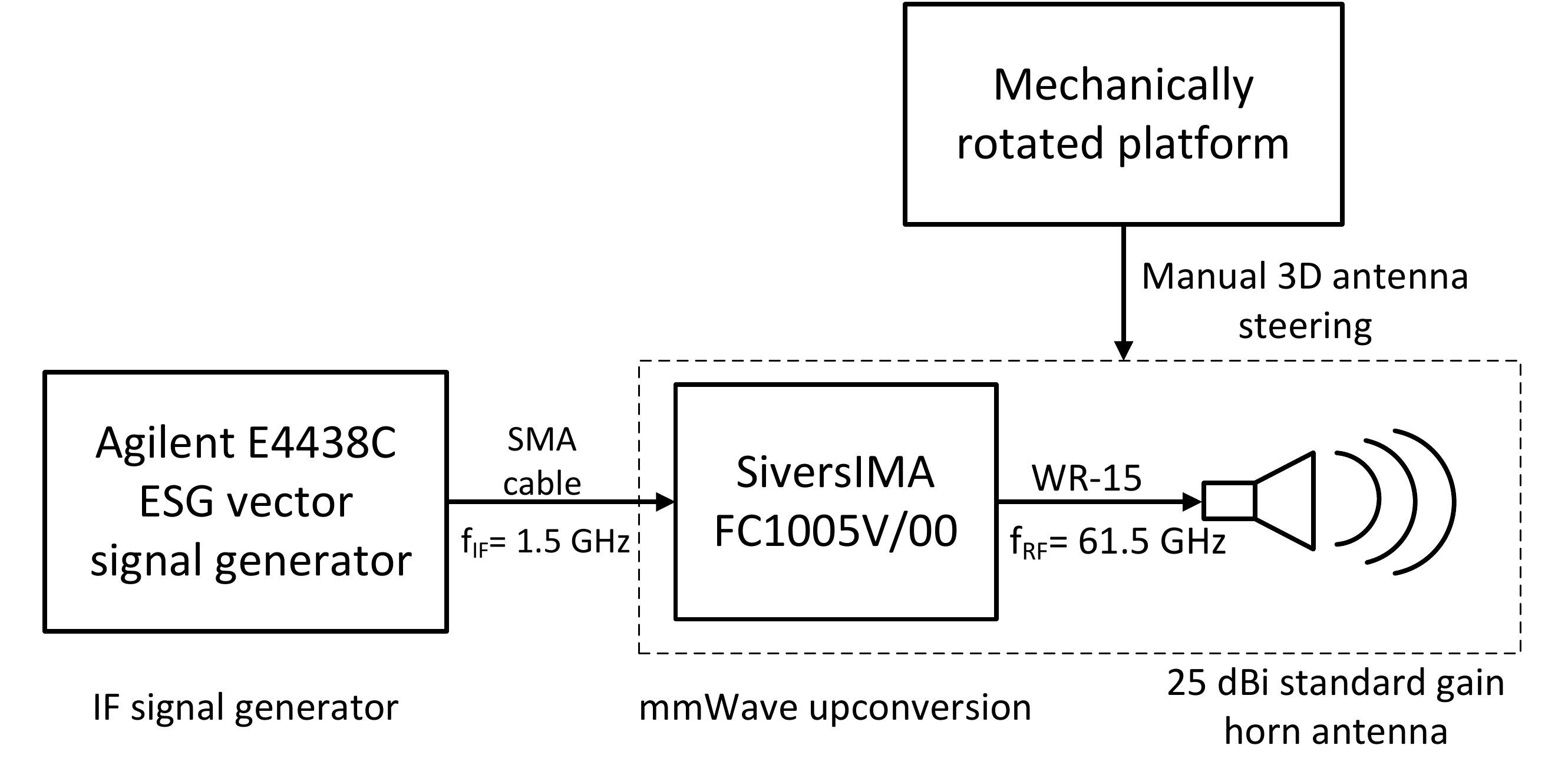}
\label{fig:SETUP_txBlock}}
\hspace{0mm}\subfloat[RX block diagram]{\includegraphics[scale=0.27]{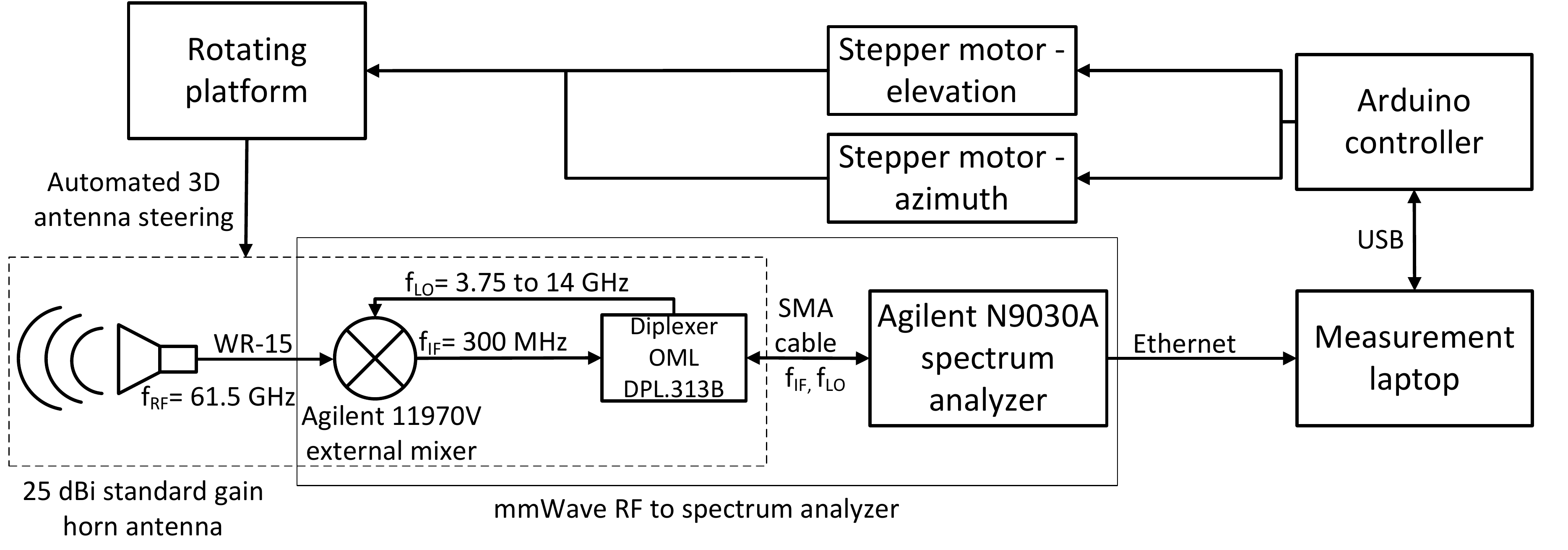}
\label{fig:SETUP_rxBlock}}\vspace{0mm}\\\subfloat[TX \emph{in situ}, looking down on RX location F]{\includegraphics[height=5cm]{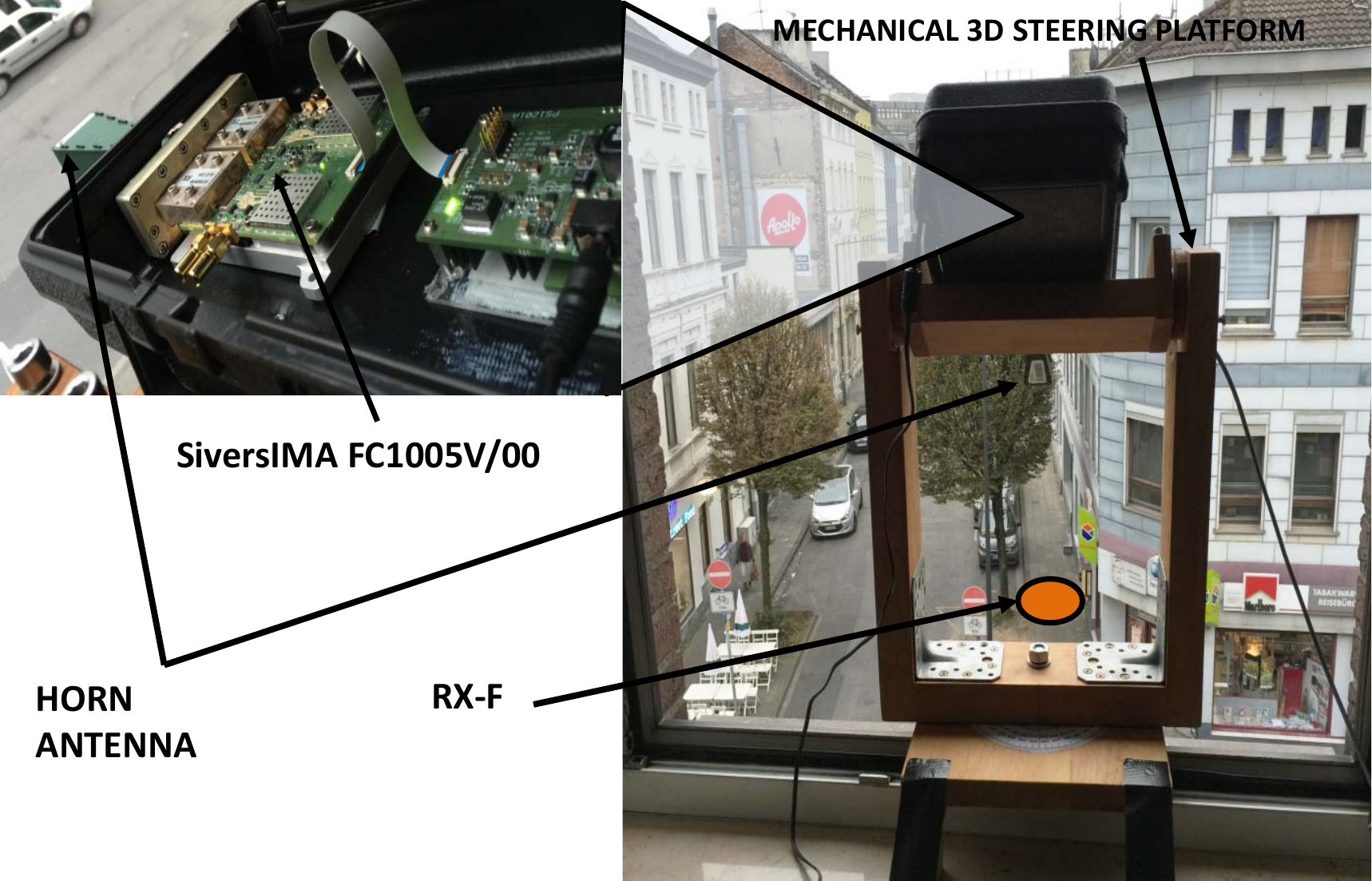}
\label{fig:SETUP_txInSitu}}
\hspace{20mm}\subfloat[RX \emph{in situ} at location A]{\includegraphics[height=5cm]{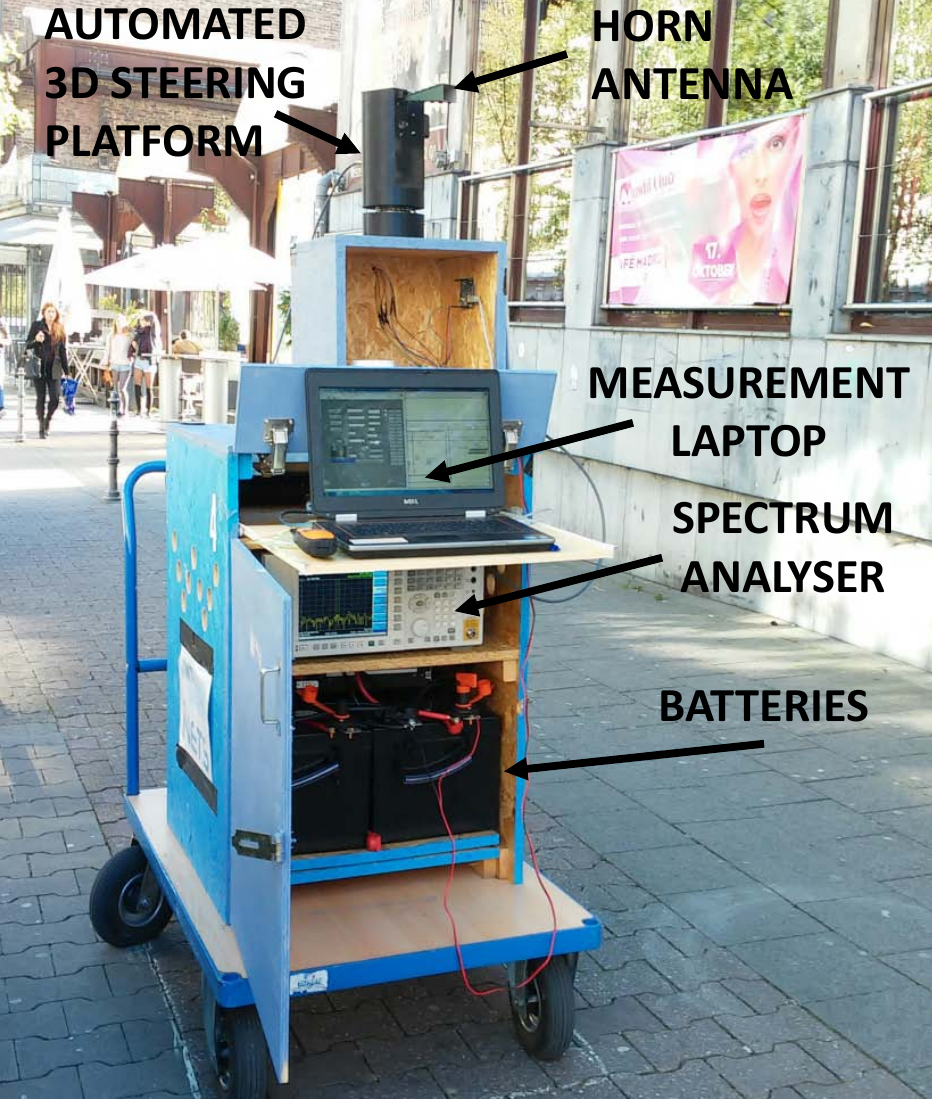}
\label{fig:SETUP_rxInSitu}}
\caption{\small{Measurement set-up, showing transmitter (TX) and receiver (RX) block diagrams and annotated \emph{in situ} photographs.}}
\label{fig:SETUP}
\end{figure*}
%....................................................................................
%....................................................................................
\begin{table}[tb!]
\begin{center}\caption{\small{TX orientations \& LOS distance per RX location}}\label{tab:Tx_el_az}
\begin{tabular}{p{0.3cm}p{0.8cm}p{2.7cm}p{1.8cm}p{0.6cm}}
\toprule
RX loc. & \# TX orient.$\dagger$ & $\phi_{tx}$  &  $\theta_{tx}$ & $d_{tx-rx}$ (3D)\\% TX-RX distance
\midrule
A & $10\times7$ & $15^{\circ}$:$5^{\circ}$:$55^{\circ}$, $95^{\circ}$& $-5^{\circ}$:$-5^{\circ}$:$-35^{\circ}$ & 28 m\\
B & $7\times4$ &  $55^{\circ}$,  $105^{\circ}$:$10^{\circ}$:$155^{\circ}$ & $-5^{\circ}$:$-10^{\circ}$:$-35^{\circ}$ & 21 m\\
C & $7\times4$ &  $55^{\circ}$:$10^{\circ}$:$115^{\circ}$ & $-5^{\circ}$:$-10^{\circ}$:$-35^{\circ}$ & 11 m\\
D & $6\times4$ &  $35^{\circ}$:$10^{\circ}$:$65^{\circ}$, $85^{\circ}$, $95^{\circ}$ & $-5^{\circ}$:$-10^{\circ}$:$-35^{\circ}$ & 19 m\\
E & $6\times4$ &   $55^{\circ}$,  $75^{\circ}$:$10^{\circ}$:$115^{\circ}$  & $-5^{\circ}$:$-10^{\circ}$:$-35^{\circ}$ & 19 m\\
F & $4\times4$ & $60^{\circ}$:$10^{\circ}$:$90^{\circ}$  & $-5^{\circ}$:$-10^{\circ}$:$-35^{\circ}$ & 24 m\\
G & $7\times4$ & $15^{\circ}$:$10^{\circ}$:$65^{\circ}$,  $95^{\circ}$  & $-5^{\circ}$:$-10^{\circ}$:$-35^{\circ}$ & 28 m\\
\bottomrule
%\multicolumn{3}{l}{$\dagger\mathrm{F}(X,Y)$: In at least $X\%$ of locations, $Y\%$ of time.}\\
\multicolumn{5}{l}{$\dagger$At each TX orientation, measurements recorded at $100\times16$ RX}\\
\multicolumn{5}{l}{orientations, $\phi_{rx}=-176.4^{\circ}$:$3.6^{\circ}$:$180^{\circ}$, $\theta_{rx}=-28.8^{\circ}$:$3.6^{\circ}$:$25.2^{\circ}$.}\\
\end{tabular}
\end{center}
\end{table}
%....................................................................................

%===========================================================================================================================
%===========================================================================================================================
\section{Measurement Setup \& Methodology}\label{SECmeasurementSetup}
Our measurement campaign was conducted over ten days in October 2015, in a busy mixed-use urban area (Fig.~\ref{fig:MAPS_all}) in the centre of the mid-sized German city of Aachen. As is typical of urban environments in many European cities, the area consists of densely built-up residential buildings of 3--5 stories, with retail outlets and restaurants on the ground floor. The transmitter was mounted at the window of a third floor apartment at a height of approximately $h_{tx}=11~\text{m}$ from the ground, overlooking a busy street and opposite an intersection with a one-way side street, as shown in Figs.~\ref{fig:MAPS_all}~and~\ref{fig:SETUP_txInSitu}. We emphasize that the transmitter height and location are representative of where a mm-wave base station might be deployed. The measurement area contains a typically heterogenous mix of building materials, e.g. shop-window glass with metal frames, metal doors, rough and smooth brickwork and concrete. This allowed us to explore a large variety of reflection opportunities from the surrounding building environment for potential NLOS transmission paths; the major reflecting materials enabling NLOS links (as observed in our measurements) are indicated in Fig.~\ref{fig:MAPS}. The receiver, shown \emph{in situ} in Fig.~\ref{fig:SETUP_rxInSitu}, was at a height of approximately $h_{rx}=1.8~\text{m}$ from the ground and placed at the seven different receiver locations~A--G in Fig.~\ref{fig:MAPS_all}. 

The future deployment of mm-wave cellular networks is preconditioned on the existence of electronically steerable on-chip phased antenna arrays. However, these are not yet available -- neither commercially, nor for system-level research experiments. Therefore, in order to emulate beamsteering of a mm-wave antenna array, in our measurements we used directional horn antennas and mechanical 3D steering platforms, whereby we set the antenna main lobe orientation by the combination of the azimuth angle $\phi$ on the horizontal plane and the elevation angle $\theta$ on the vertical plane. A unique pair of transmitter and receiver antenna orientation may thus be expressed as the 4-tuple $(\phi_{tx}, \theta_{tx}, \phi_{rx}, \theta_{rx})$.  We steered the transmit antenna manually using the platform shown in Fig.~\ref{fig:SETUP_txInSitu}, whereas at the receiver we used the custom 3D-printed platform housing the antenna shown in Fig.~\ref{fig:SETUP_rxInSitu} with a controller and stepper motors  ($3.6^{\circ}$ resolution) to automatically step through receiver orientations.

At each individual receiver location, measurements were conducted for multiple (azimuth~$\times$~elevation) combinations of transmitter antenna orientation\footnote{We considered only TX angles which may result in a LOS or reflected NLOS link, given the geometry of the urban environment at each RX location; e.g. for RX G, $65^{\circ}<\phi_{tx}<95^{\circ}$ was omitted as no reflected NLOS path is feasible if the TX is pointing down the side street.}, as specified in Table~\ref{tab:Tx_el_az}. In turn, for each considered transmitter orientation, we measured the received signal strength (RSS) for $100\times16$ (azimuth~$\times$~elevation) combinations of receiver antenna orientation. The angular granularity of our measurement sweeps is thus significantly higher than earlier works; e.g. the $28~\text{GHz}$ AoA measurement results in~\cite{Samimi2013_28GHz_AoA} were based on sweeps over only $(3\times1\times36\times3)$ combinations of $(\phi_{tx}, \theta_{tx}, \phi_{rx}, \theta_{rx})$, whereas the $60~\text{GHz}$ measurements in~\cite{Rappaport2012_3860GHz} searched over the azimuth and elevation planes for substantial signals randomly rather than systematically. We note that each of our single complete 3D receiver antenna sweeps takes approximately 15 minutes, which represents a significant measurement time-budget challenge, given all combinations of considered transmitter orientations and receiver locations. Consequently, we chose the granularity of transmitter antenna steering to be comparable to the antenna beamwidth of $10^{\circ}$, whereas the finer granularity of steering at the receiver was chosen to collect detailed angular link measurements with respect to the surrounding urban building structures and materials, thus yielding insight into the beamsteering requirements of urban mm-wave cellular networks.
%for and a fixed elevation angle and three azimuth angles at the transmitter

Figs.~\ref{fig:SETUP_txBlock}~and~\ref{fig:SETUP_rxBlock} detail our mm-wave measurement setup at the transmitter and receiver side, respectively. At the transmitter, Agilent E4438C ESG vector signal generator is used to generate a continuous wave signal\footnote{We emphasize that we use narrowband power measurements without loss of generality, as we are interested only in measuring expected RSS (i.e. path loss) for a given TX/RX orientation, and not in time-domain characterization of the mm-wave channel as in the channel sounding measurements of~\cite{Rappaport2015_MeasurmentsSurvey, FraunhoferBerlin2014}.} of $P_{tx,IF}=-18~\text{dBm}$ at the intermediate frequency of $f_{IF}=1.5~\text{GHz}$, which is then fed via a low-loss SMA cable to the SiversIMA FC1005V/100~\cite{siversima} mm-wave upconverter. The output of the mm-wave upconverter is a $P_{tx,RF}=5~\text{dBm}$ signal at the centre frequency of $f_{RF}=61.5~\text{GHz}$, which is finally fed via the WR-15 waveguide interface to a $G_{tx}=25~\text{dBi}$ standard gain horn antenna with a half power beamwidth of approximately $10^{\circ}$~\cite{flann_antenna}. The equivalent isotropically radiated power (EIRP) of our setup is thus $30~\text{dBm}$, in line with US and EU spectrum regulation~\cite{fcc, europe}. At the receiver, the RF signal received by an identical $G_{rx}=25~\text{dBi}$ standard gain horn antenna is fed into the Agilent N9030A spectrum analyzer (SA) which records the received signal power. We note that Agilent N9030A SA natively supports RF input only up to $26.5~\text{GHz}$, but using the Agilent 11970V external mixer (in conjunction with the OML DPL.313B diplexer) enables measuring power of RF signals in the range of $50-75~\text{GHz}$. For each considered $(\phi_{tx}, \theta_{tx}, \phi_{rx}, \theta_{rx})$ antenna orientation combination at a given receiver location, the received power was measured for $1~\text{ms}$ in zero-span mode, whereby the SA records samples in the time domain, for the frequency bin of resolution bandwidth\footnote{We chose this RBW after observing during calibration tests that the original IF continuous wave signal at the transmitter is broadened somewhat after up-conversion to mm-wave RF.} $RBW=40~\text{kHz}$, centred at $61.5~\text{GHz}$. The noise floor\footnote{We can thus measure a maximum path loss of $139~\text{dB}$; this is somewhat less sensitive than the measurement setup of Rappaport~\emph{et~al.}~\cite{Rappaport2015_MeasurmentsSurvey}, but  more than sufficient for our purposes of discerning feasible high data rate links.} of our overall receiver setup (i.e. taking into account thermal noise and the losses at the SA, external mixer, and diplexer) is $-84~\text{dBm}$.% Finally, the power measurements from the SA are recorded on the measurement laptop, which also controls the automated 3D steering of the receive antenna via a LabView interface to the Arduino controller and stepper motors. 
%, with significant foot traffic on the footpath and vehicle movement on the road.
%\FloatBarrier
%, given e.g. the minimum receiver sensitivity of $-75.5~\text{dBm}$ for LTE with $1~\text{GHz}$ channel bandwidth.
%After performing calibration tests to investigate the stability of our setup over time and under various temperature and humidity conditions, we observed that the original IF continuous wave signal at the transmitter is broadened somewhat after up-conversion to mm-wave RF; we thus chose to use a RBW of $40~\text{kHz}$ at the receiver.

%===========================================================================================================================
%\clearpage
\section{Measurement Results \& Analysis}\label{SECresults}
\begin{figure}[!tb]
\centering
\subfloat[RX location A]{\includegraphics[width=\columnwidth]{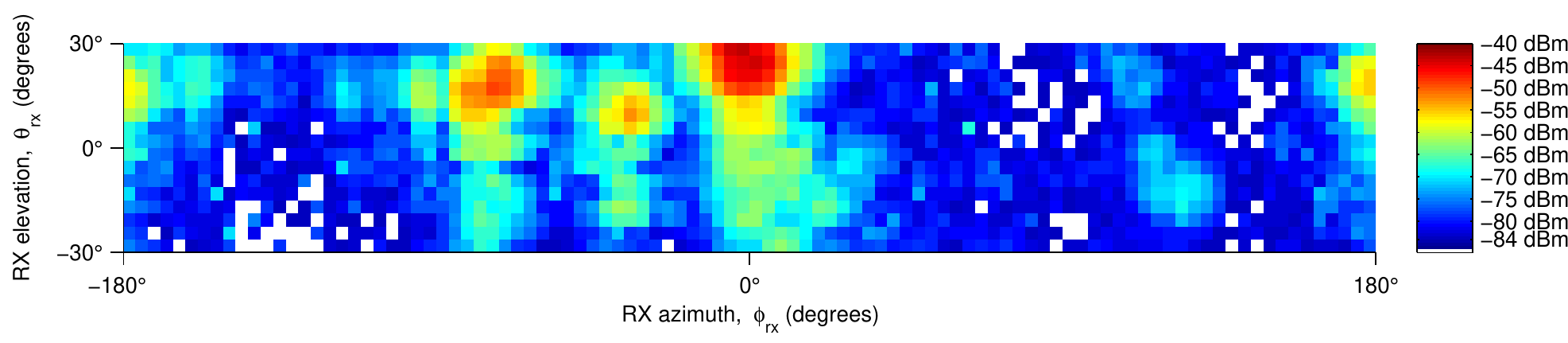}
\label{fig:A}}
\\\subfloat[RX location B]{\includegraphics[width=\columnwidth]{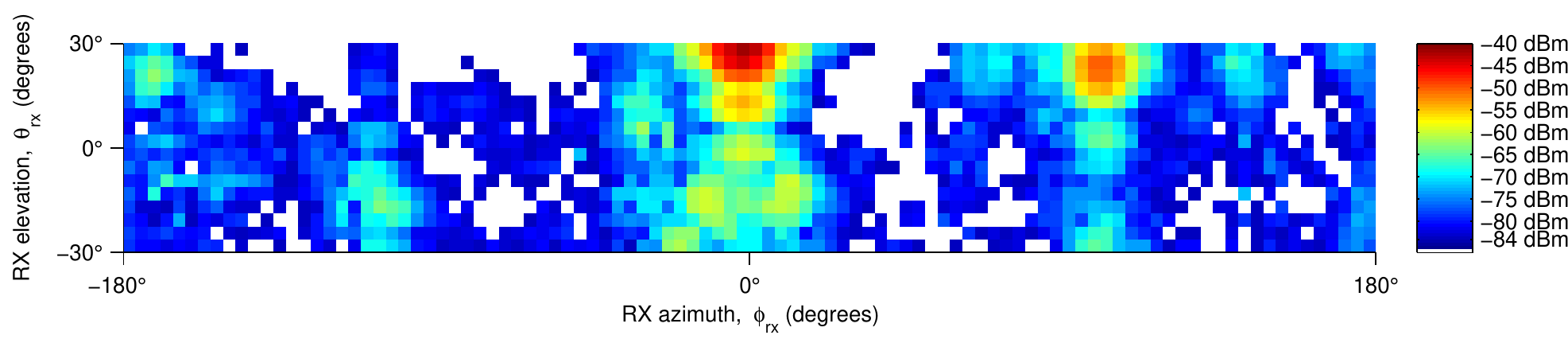}
\label{fig:A}}
\\\subfloat[RX location C]{\includegraphics[width=\columnwidth]{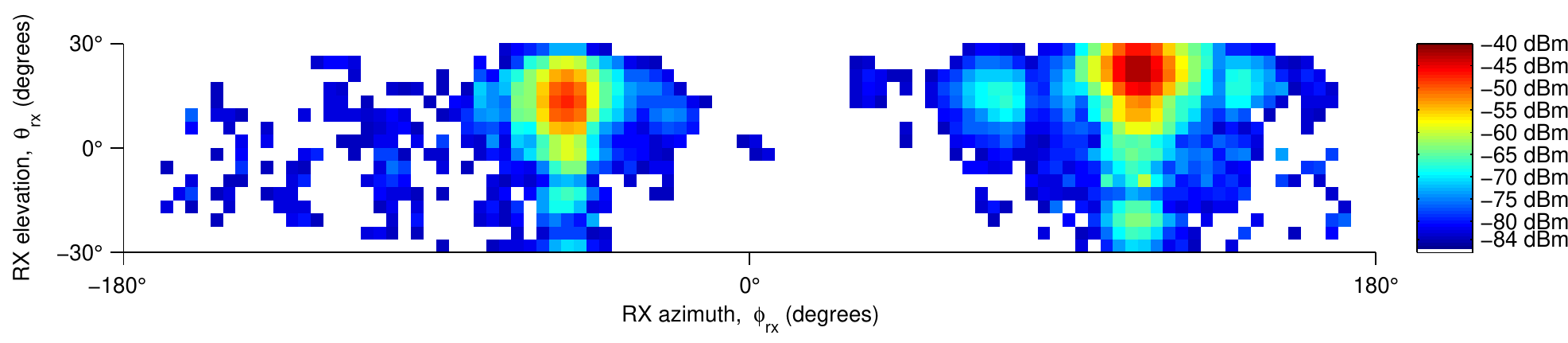}
\label{fig:A}}
\\\subfloat[RX location D]{\includegraphics[width=\columnwidth]{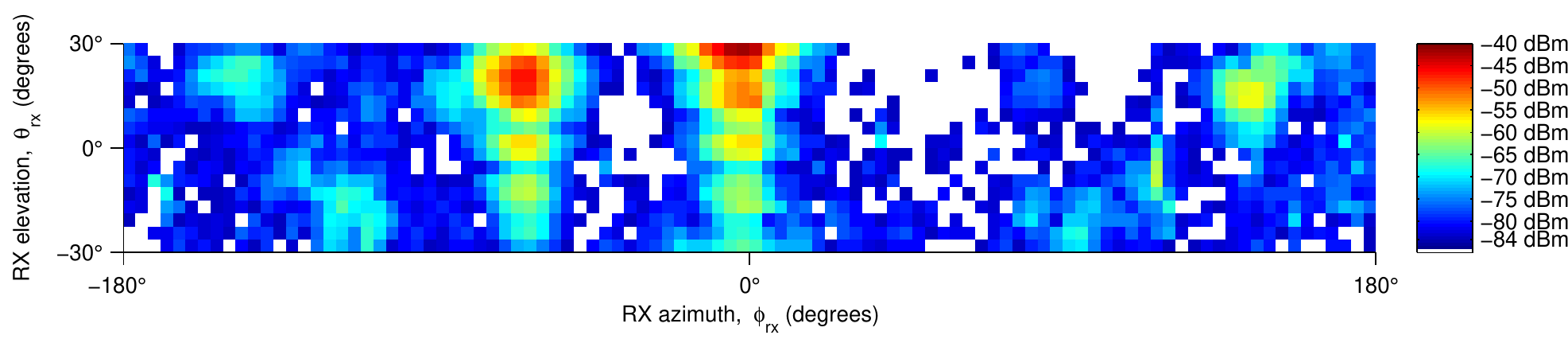}
\label{fig:A}}
\\\subfloat[RX location E]{\includegraphics[width=\columnwidth]{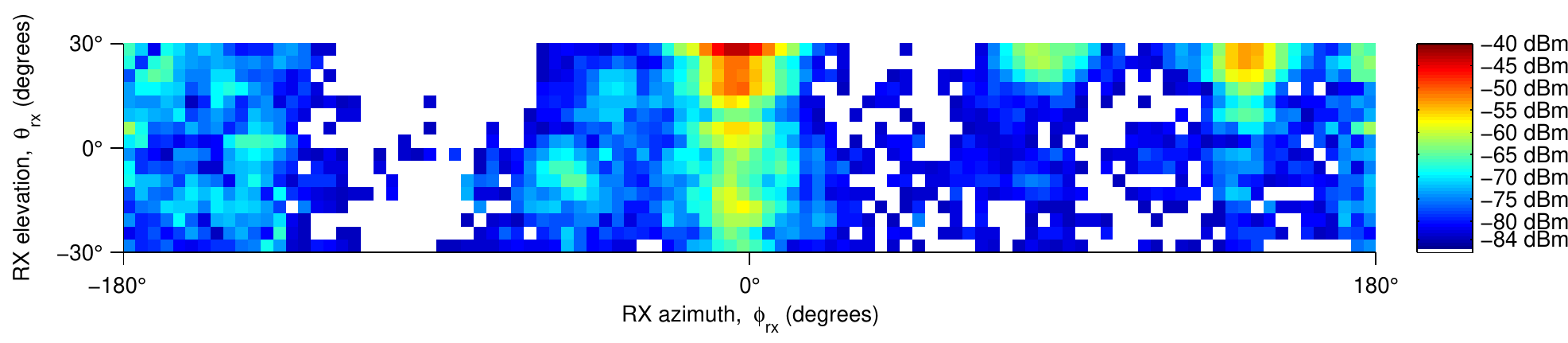}
\label{fig:A}}
\\\subfloat[RX location F]{\includegraphics[width=\columnwidth]{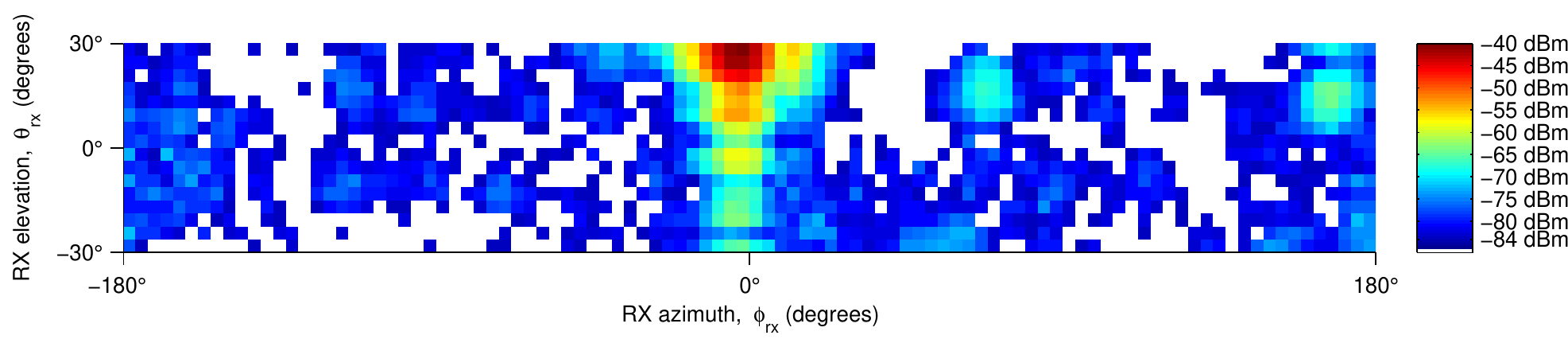}
\label{fig:A}}
\\\subfloat[RX location G]{\includegraphics[width=\columnwidth]{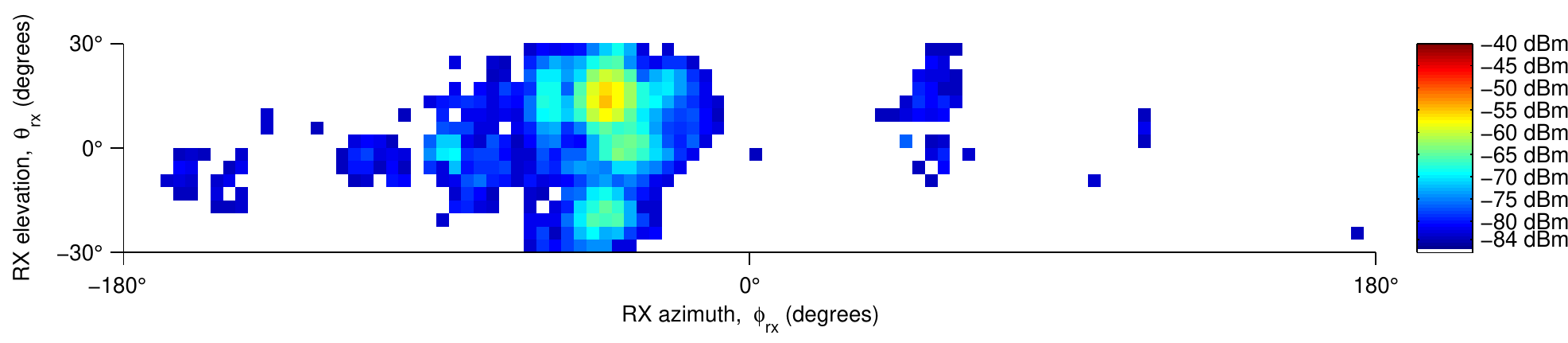}
\label{fig:A}}
\caption{\small{Heatmap of RSS (dBm) measured over all 100~$\times$~16 (azimuth $\times$ elevation) receiver antenna orientations (highest RSS over all transmitter orientations in Table~\ref{tab:Tx_el_az}) for each receiver location (measurement setup noise floor is $-84~\text{dBm}$).}}
\label{fig:heatmapRX_RSSdBm}
\end{figure}

The heatmaps in Fig.~\ref{fig:heatmapRX_RSSdBm}, per receiver location, show for each receive antenna orientation\footnote{We assume $\theta_{rx}=0^{\circ}$ is the horizon and $\phi_{rx}=0^{\circ}$ is the geometric LOS direction to the TX for each RX location.} $(\phi_{rx}, \theta_{rx})$ the highest RSS measured over all transmit antenna orientations $(\phi_{tx}, \theta_{tx})$ considered at that receiver location. We note these results are thus optimistic, in the sense that they imply optimal transmitter alignment for any given receiver orientation\footnote{For the sake of brevity, throughout this paper we must focus on results for the best measured transmitter orientation and defer analysis of the combined beamsteering effort at the receiver and transmitter to a subsequent publication.}.

Strong LOS clusters are evident in Fig.~\ref{fig:heatmapRX_RSSdBm} at receiver locations A, B, D, E, and F. Up to four strong NLOS clusters per location were also observed, largely corresponding to single-bounce reflections from glass windows, as illustrated in Fig.~\ref{fig:MAPS}. A rich inventory of alternative NLOS paths is crucial for mm-wave networks, as it makes feasible switching to another transmission path if the strongest link direction becomes blocked by moving obstacles. Somewhat surprisingly, the average number of transmission paths we observed per receiver location is consistent with that found in New York City at $28~\text{GHz}$ in~\cite{Samimi2013_28GHz_AoA}. We note though, that several of our observed NLOS paths were due to reflections from parked cars rather than buildings, as shown in Fig.~\ref{fig:MAPS}; such NLOS paths are expected to be transient and thus only useful for ``opportunistic'' beamsteering. Nonetheless, the existence of several major NLOS clusters, even in non-metropolitan environments as exemplified by our study area, is encouraging for mm-wave pico-cellular urban deployments.

However, our results also demonstrate the strongly site-specific nature of mm-wave connectivity. Fig.~\ref{fig:heatmapRX_RSSdBm} shows, for example, that we received a signal above our noise floor of $-84~\text{dBm}$ for almost $95\%$ of considered receiver orientations at location A, but for under $25\%$ at location G. Namely, the available degrees of freedom for beamsteering highly varies even \emph{within} our small pico-cellular site (as shown in Table~\ref{tab:Tx_el_az}, our effective cell radius is under $30~\text{m}$), according to the placement of the receiver relative to local reflecting surfaces. This highlights the importance of closely considering the geometry and building materials of the specific urban environment at a potential cell site in mm-wave network planning. A related potential implication is that more individualistic and complex per-site configuration may be required for installing mm-wave deployments than in traditional cellular networks.

%and suggests that more detailed and individualistic per-site planning may be 

%The (the available degrees of freedom for beamsteering) number of receiver orientations where a signal strength above noise floor was observed/measured depends highly on the relative location of the receiver, and the corresponding availability of LOS or reflected NLOS links. 
%---------------------------------------------------------------------------------------------------------------------------
%===========================================================================================================================
%%....................................................................................
%....................................................................................

%%....................................................................................
\begin{figure}[!tb]
\centering
\subfloat[RX location A]{\includegraphics[width=\columnwidth]{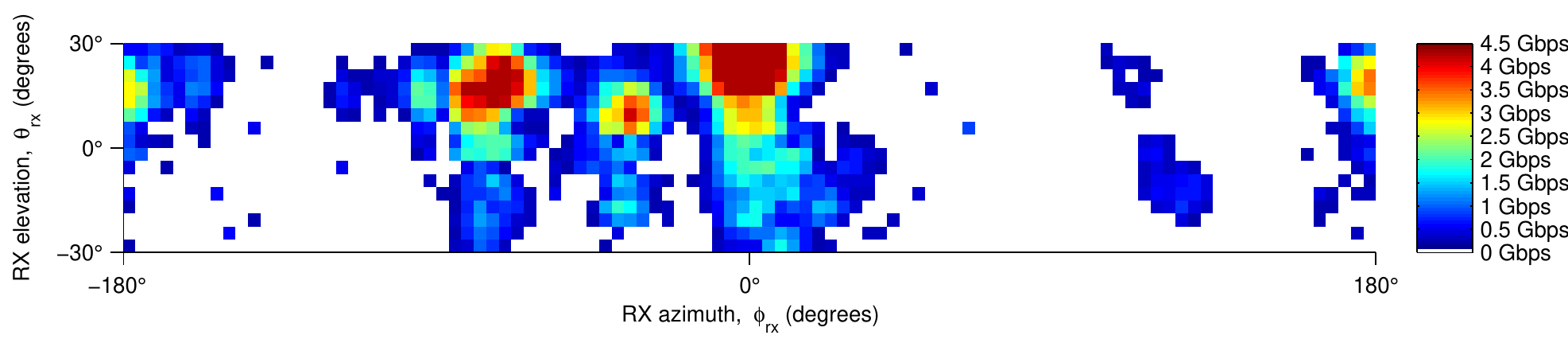}
\label{fig:A}}
\\\subfloat[RX location B]{\includegraphics[width=\columnwidth]{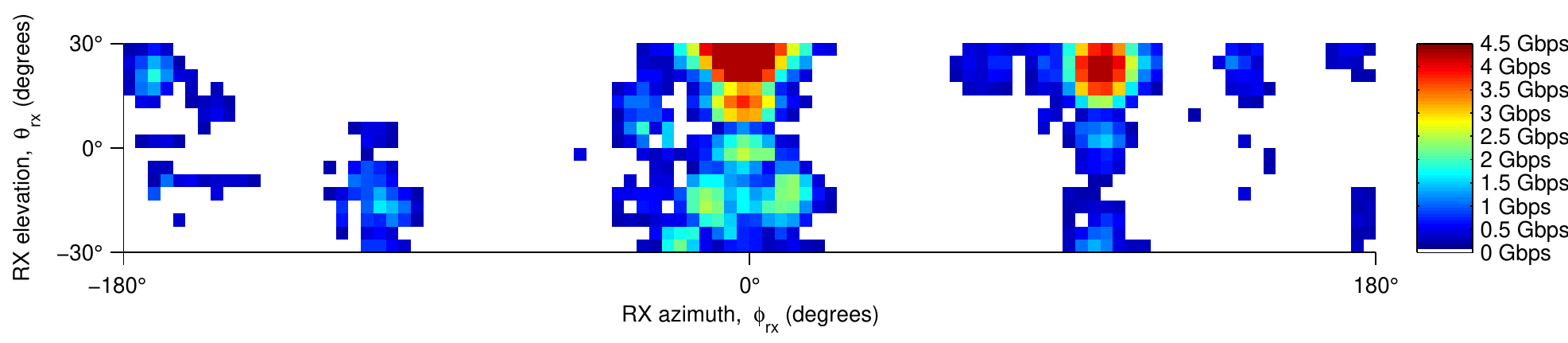}
\label{fig:A}}
\\\subfloat[RX location C]{\includegraphics[width=\columnwidth]{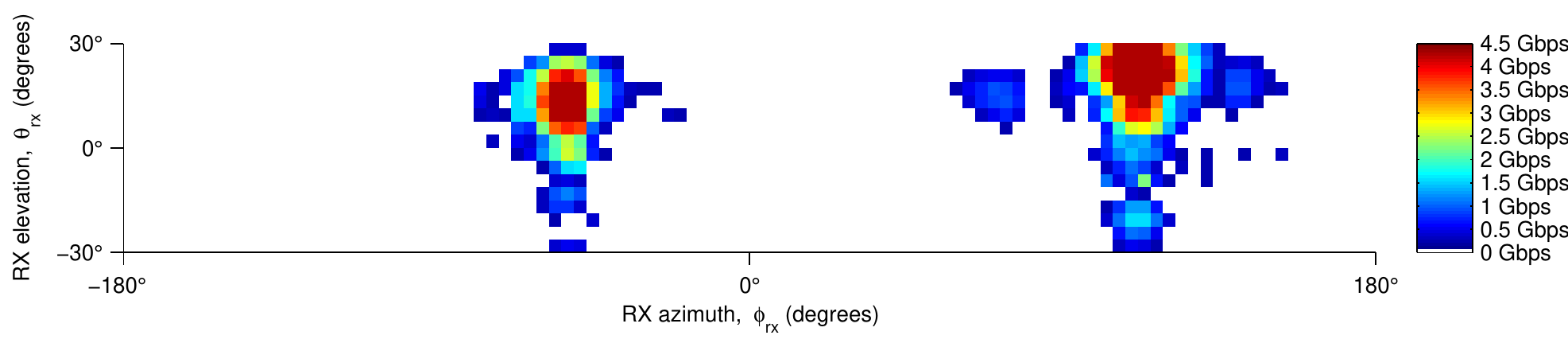}
\label{fig:A}}
\\\subfloat[RX location D]{\includegraphics[width=\columnwidth]{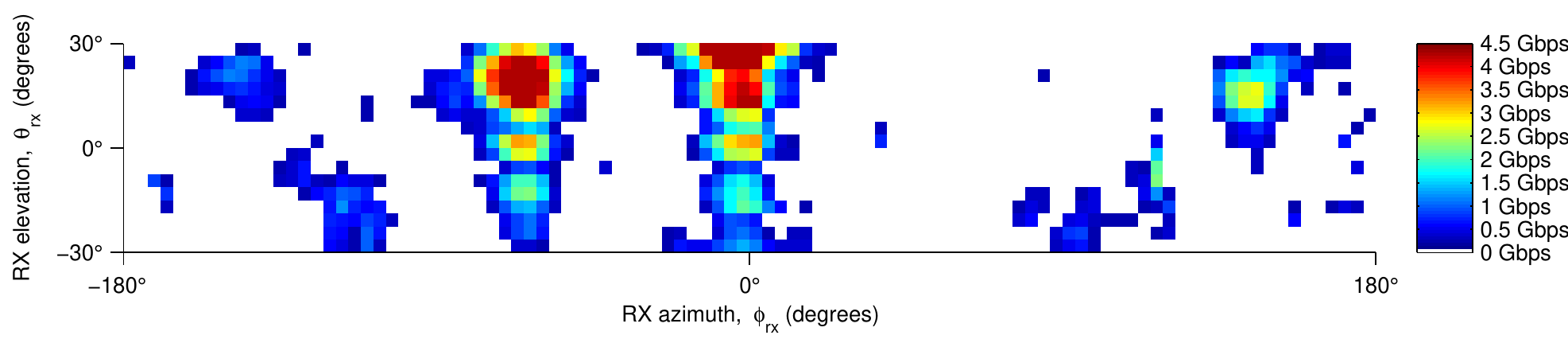}
\label{fig:A}}
\\\subfloat[RX location E]{\includegraphics[width=\columnwidth]{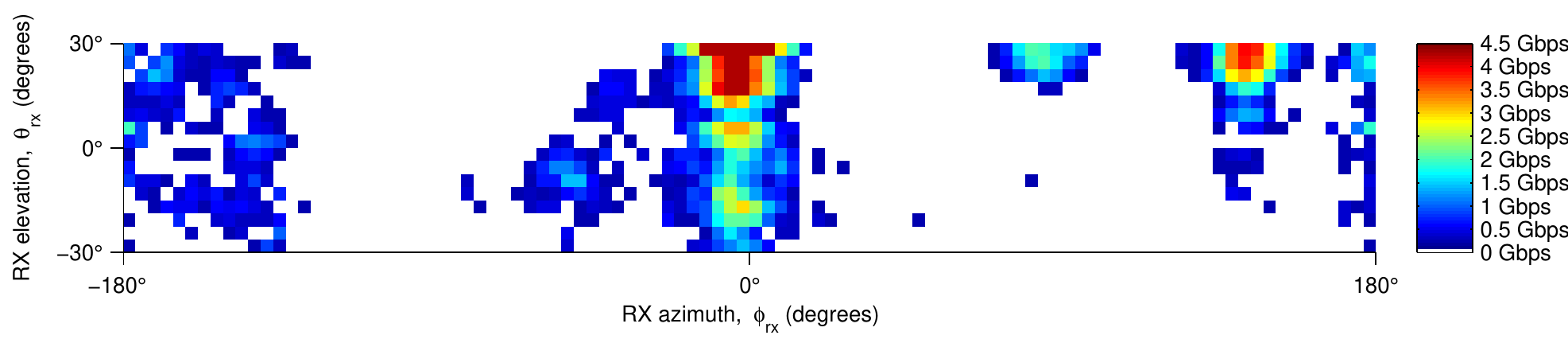}
\label{fig:A}}
\\\subfloat[RX location F]{\includegraphics[width=\columnwidth]{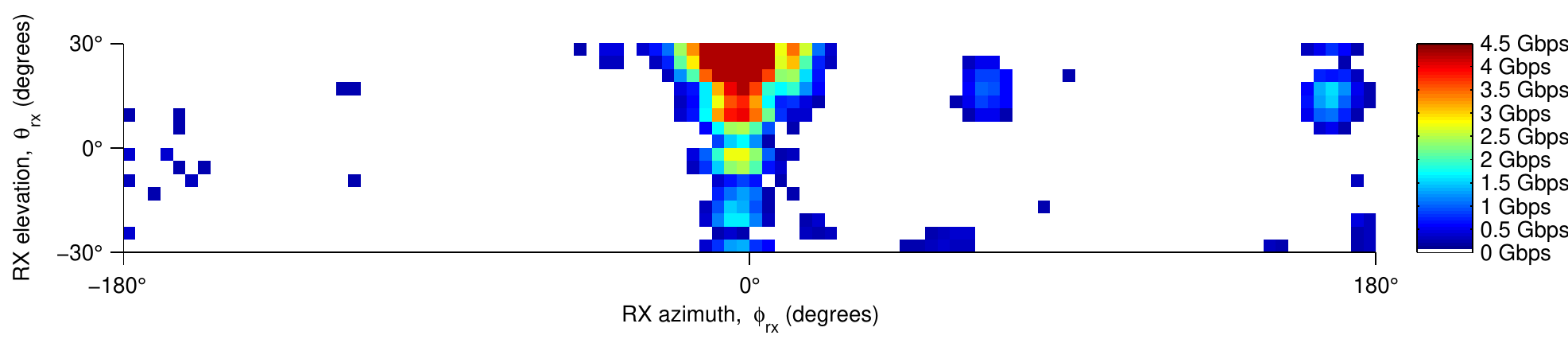}
\label{fig:A}}
\\\subfloat[RX location G]{\includegraphics[width=\columnwidth]{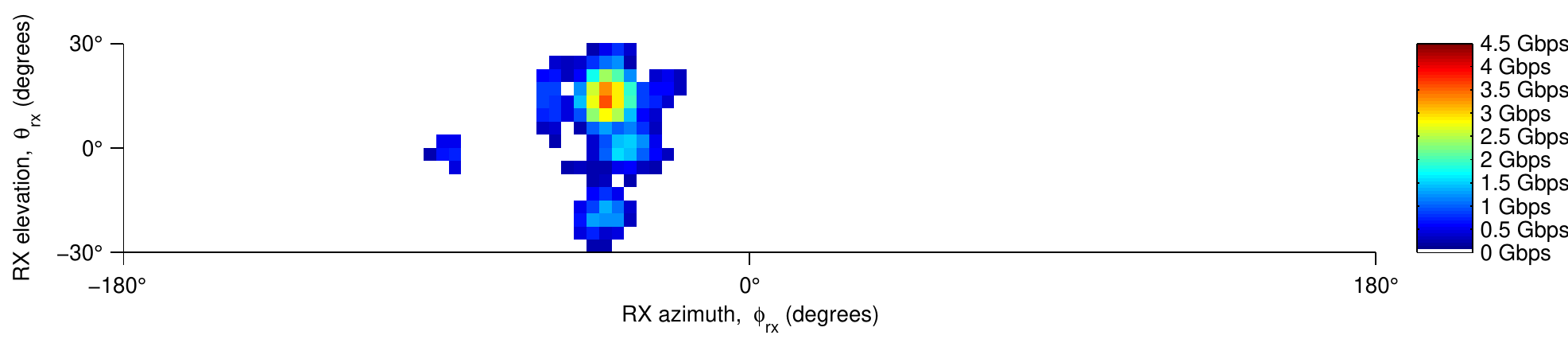}
\label{fig:A}}
\caption{\small{Heatmap of achievable data rate over all 100~$\times$~16 (azimuth $\times$ elevation) measured receiver antenna orientations (maximum over all transmitter orientations in Table~\ref{tab:Tx_el_az}), assuming 1 GHz channel bandwidth and LTE autorate function (i.e. -75.5 dBm minimum receiver sensitivity).}}
\label{fig:heatmapRX_DataRateGbps}
\end{figure}
\begin{figure}[!tb]%!tb
\centering
\subfloat[RX location A]{\includegraphics[scale=0.44]{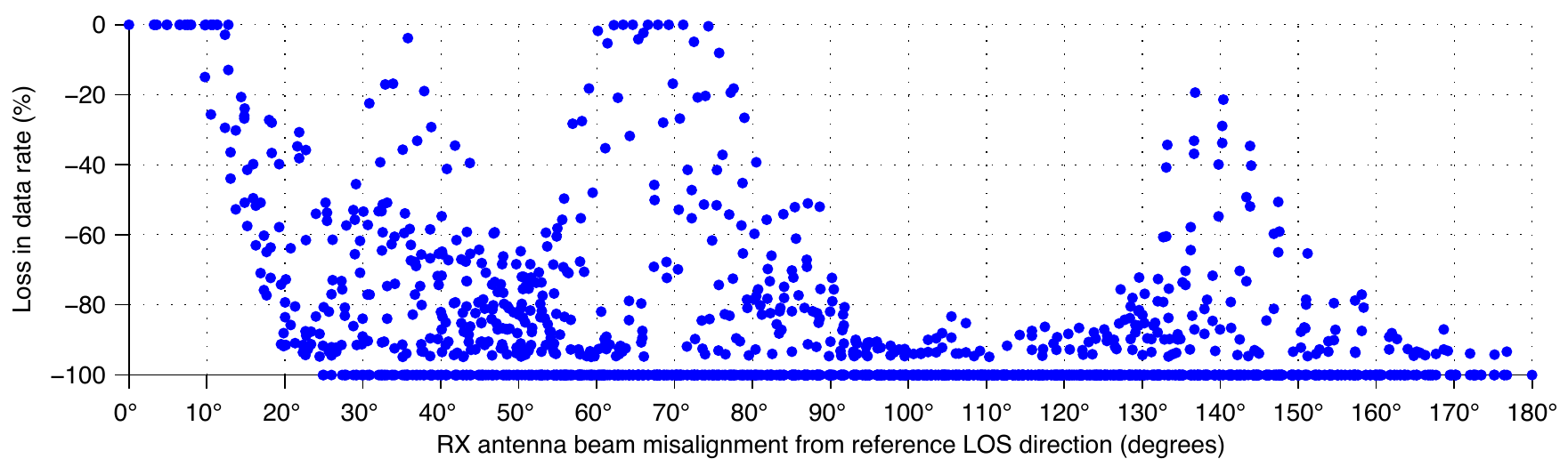}
\label{fig:scatterRX_beamMisalignment_A}}
\\\subfloat[RX location B]{\includegraphics[scale=0.44]{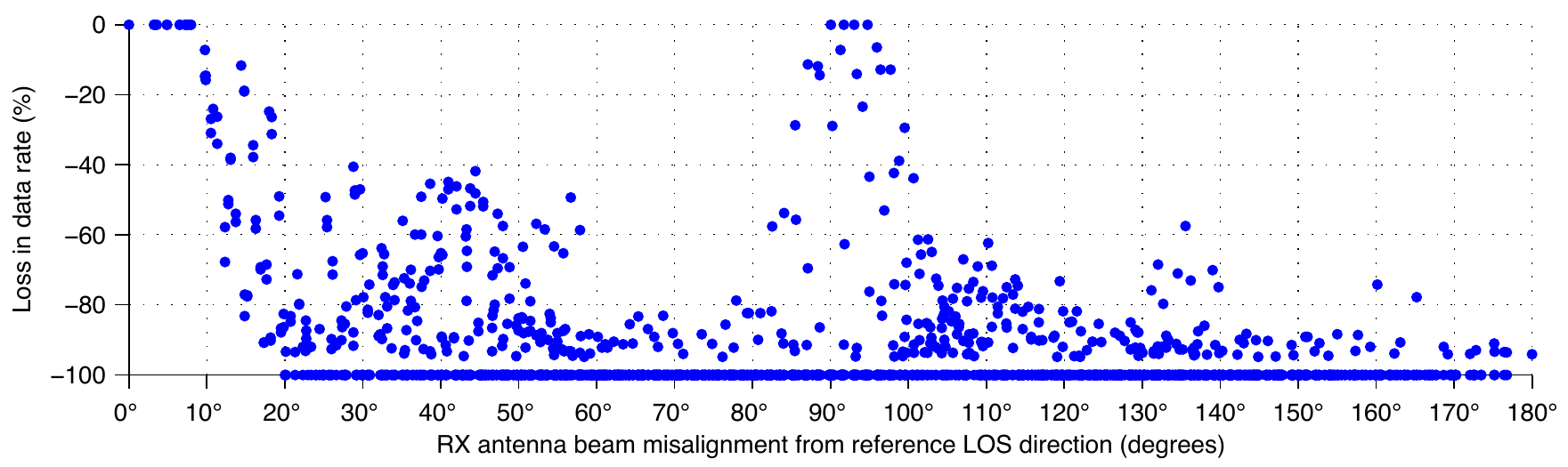}
\label{fig:scatterRX_beamMisalignment_B}}
\\\subfloat[RX location C]{\includegraphics[scale=0.44]{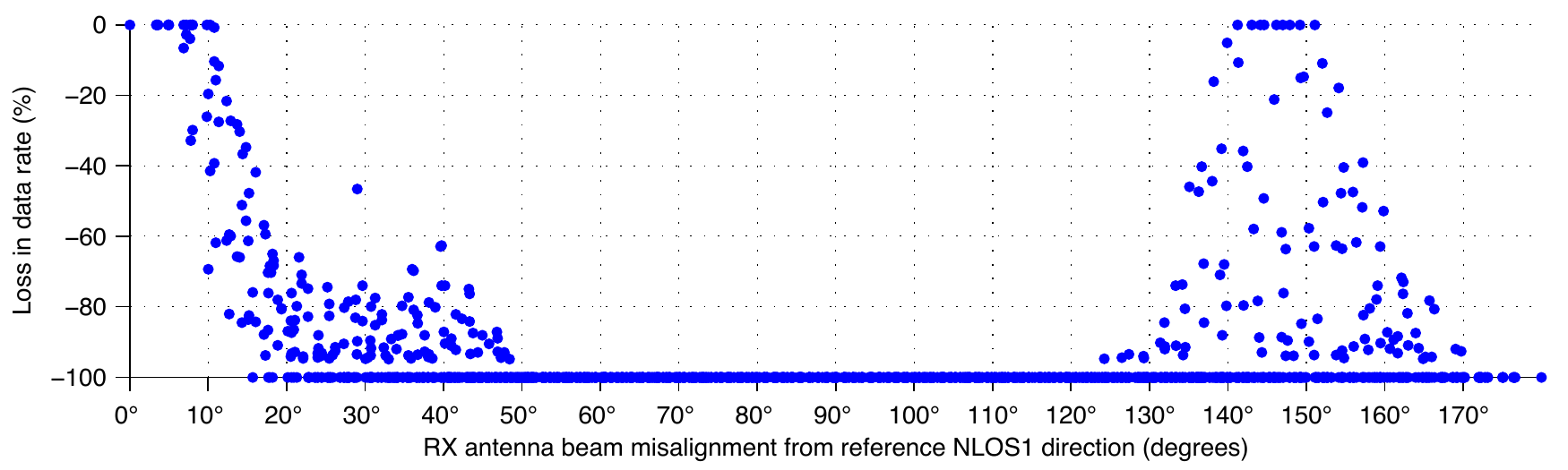}
\label{fig:scatterRX_beamMisalignment_C}}
\\\subfloat[RX location D]{\includegraphics[scale=0.44]{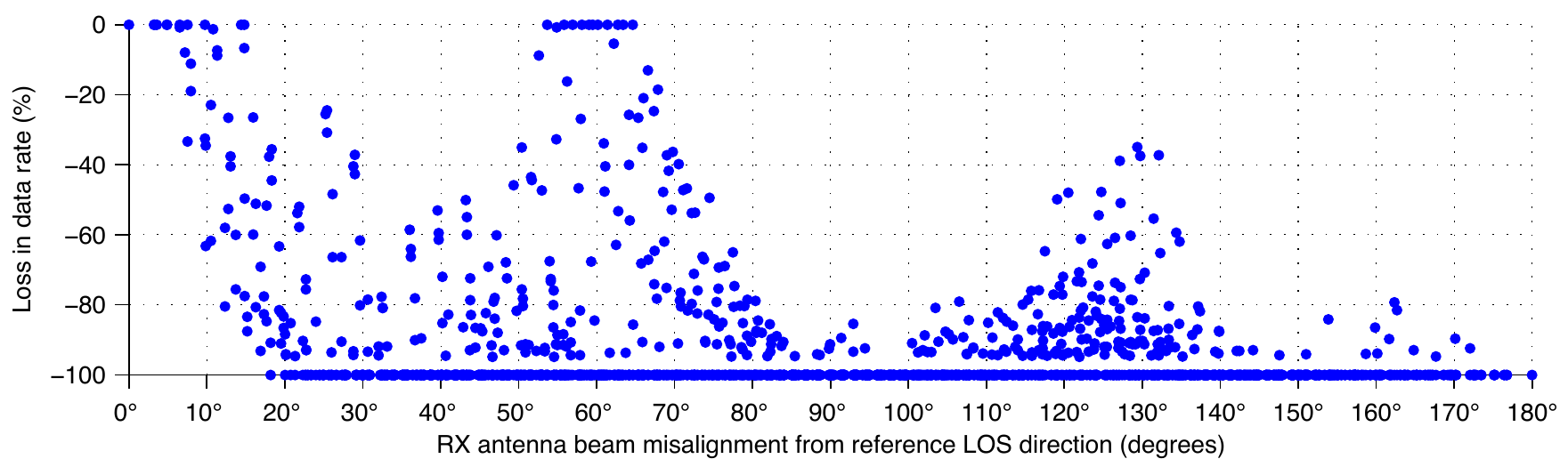}
\label{fig:scatterRX_beamMisalignment_E}}
\\\subfloat[RX location E]{\includegraphics[scale=0.44]{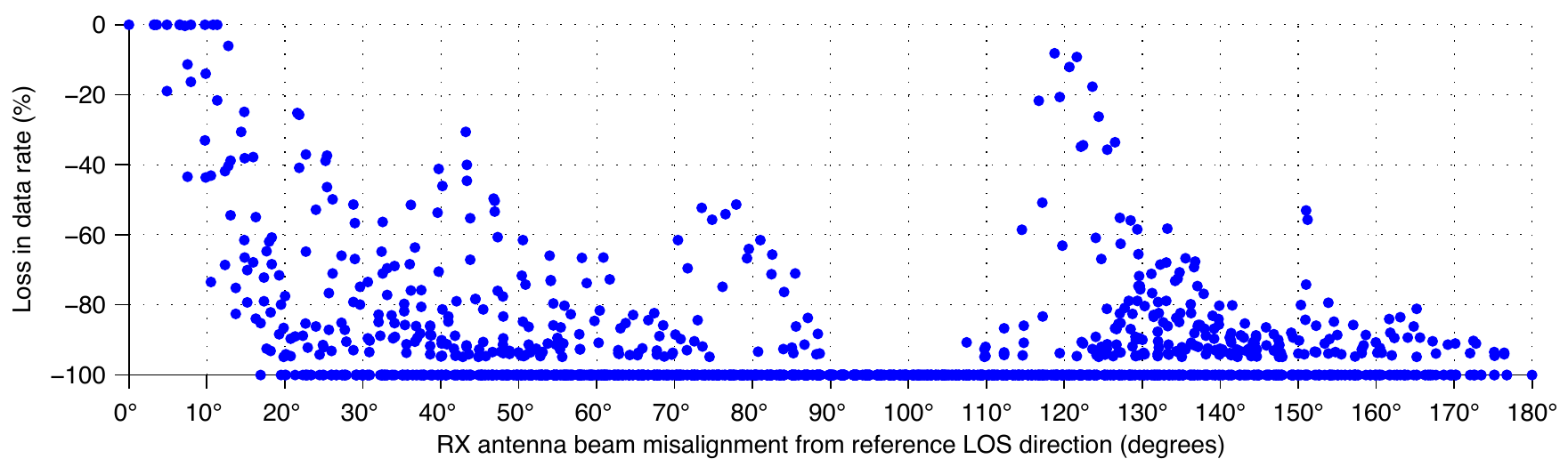}
\label{fig:scatterRX_beamMisalignment_E}}
\\\subfloat[RX location F]{\includegraphics[scale=0.44]{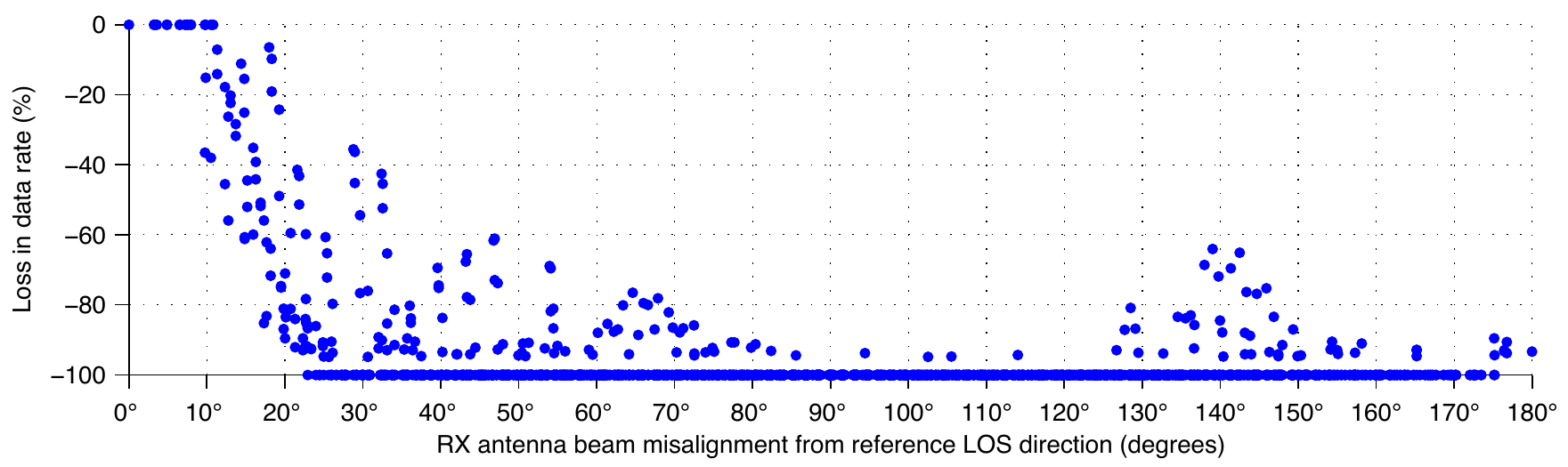}
\label{fig:scatterRX_beamMisalignment_F}}
\\\subfloat[RX location G]{\includegraphics[scale=0.44]{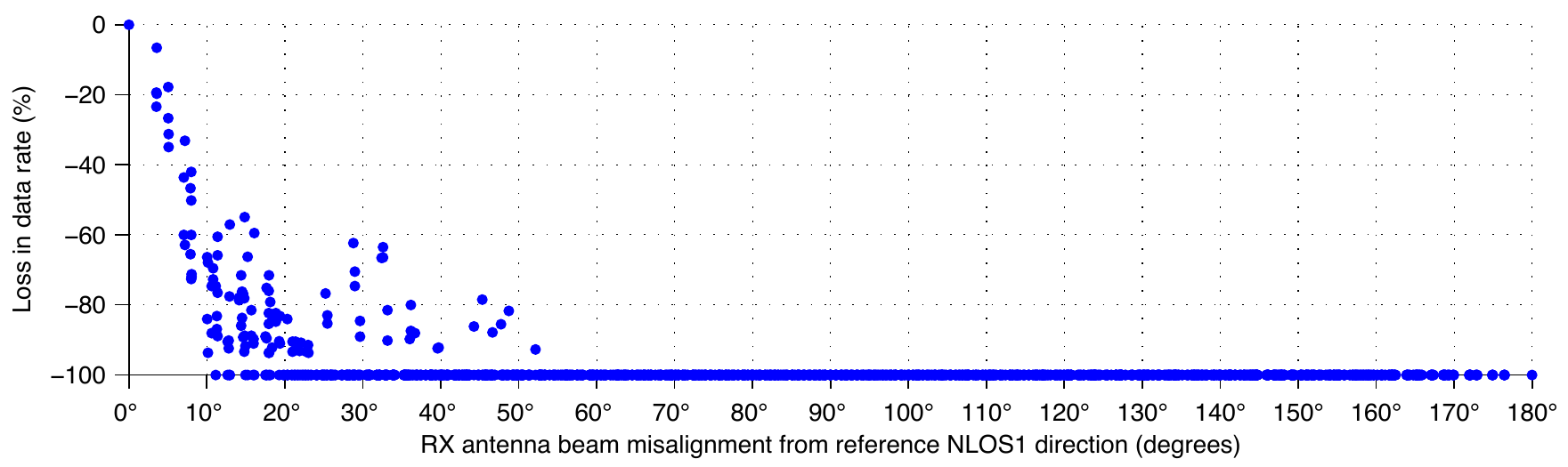}
\label{fig:scatterRX_beamMisalignment_G}}
\caption{\small{Percent loss in data rate due to receiver antenna beam misalignment with respect to the strongest link direction (assuming best transmitter antenna orientation for each receiver orientation).}}
\label{fig:scatterRX_beamMisalignment}
\end{figure}
%....................................................................................

%%%....................................................................................
%\begin{figure}[!tb]
%\centering
%\includegraphics[width=0.9\columnwidth]{CDF_DATA_RATE_Gbps_perPos2}
%\caption{\small{Distribution of the achievable data rate measured over all 16~~$\times$~100 (elevation $\times$ azimuth) receiver antenna orientations for each receiver (RX) position, assuming LTE autorate function and optimal TX antenna orientations.}}
%\label{fig:ABC}
%\end{figure}
%%....................................................................................

In order to interpret the practical utility of the measured RSS values in Fig.~\ref{fig:heatmapRX_RSSdBm} for a mm-wave cellular network deployment, we map them to an estimate of the achievable data rate in  Fig.~\ref{fig:heatmapRX_DataRateGbps}. We assume a $1~\text{GHz}$ channel bandwidth, a receiver noise figure of $10~\text{dB}$, implementation margin of $5~\text{dB}$, and the LTE Rel.~12 autorate function given in~\cite{3gppLTE} as the truncated Shannon bound with attenuation factor $\alpha=0.75$, minimum SNR of $-6.5~\text{dB}$, and maximum SNR of $17~\text{dB}$. The corresponding top achievable data rate is $4.2568~\text{Gbps}$, whereas the minimum receiver sensitivity is $-75.5~\text{dBm}$ (for a data rate of $218.6~\text{Mbps}$).

Fig.~\ref{fig:heatmapRX_DataRateGbps} reveals the far more restricted range of valid antenna orientations for obtaining high-speed cellular mm-wave links: only under a third of the antenna orientations with RSS above the noise floor in Fig.~\ref{fig:heatmapRX_RSSdBm} result in a valid LTE-like link in Fig.~\ref{fig:heatmapRX_DataRateGbps}. Moreover, on average only $9\%$ of all considered antenna orientations per receiver location result in a link with data rate of over $1~\text{Gbps}$, whereas the top data rate of $4.2568~\text{Gbps}$ is achievable for no more than $1\%$ of the antenna orientations on average. These results reveal that very stringent beamsteering requirements must be met if mm-wave cellular networks are to deliver the promise of multi-Gbps connectivity. 

To analyze in more detail the sensitivity of mm-wave outdoor cellular links to suboptimal beamsteering, in Fig.~\ref{fig:scatterRX_beamMisalignment} we consider the degradation in data rate due to receive antenna beam misalignment with respect to the strongest link direction. Specifically, the scatter plots in Fig.~\ref{fig:scatterRX_beamMisalignment} were obtained by computing, for each of the $1600$ antenna orientations at a given receiver location, the angular distance from the centre of the LOS cluster (or strongest NLOS for locations C and G) and the corresponding data rate drop in Fig.~\ref{fig:heatmapRX_DataRateGbps} from this reference optimal orientation. The clusters corresponding to the valid LOS/NLOS links in Fig.~\ref{fig:heatmapRX_DataRateGbps} are thus evident as peaks in Fig.~\ref{fig:scatterRX_beamMisalignment}.

Let us first consider how sensitive the strongest LOS/NLOS links, corresponding to the leftmost cluster-peaks in Fig.~\ref{fig:scatterRX_beamMisalignment} are. A misalignment of around $10^{\circ}$ -- in the order of the antenna beamwidth -- results in almost no reduction in data rate for LOS links at locations A, B, and F. However, at the remaining receiver locations, the strongest LOS/NLOS links  typically suffer a data rate drop of $20\%$ due to a $10^{\circ}$ misalignment and an average drop of $50-100\%$ due to a $20^{\circ}$ misalignment. The weaker NLOS links are even more sensitive to misalignment, e.g. as evident from the second cluster-peak in Fig.~\ref{fig:scatterRX_beamMisalignment_F} at the angular distance of $139^{\circ}$ from the reference LOS orientation. This poses a great challenge to providing stable high-speed connectivity to mobile users in a mm-wave cellular network, given that changes in orientation in the order of $10^{\circ}$ would occur very frequently even for a quasi-stationary user, i.e. not walking but holding the phone in their hand.

The receive antenna beam would need to be continuously re-steered to compensate for such ``micro-movements'', and to provide the requisite channel state information, the user orientation relative to the environment should be continuously monitored. Furthermore, to support user mobility and overcome blockage of preferred transmission paths by moving obstacles, the transmit antenna would also need to be frequently re-steered (as illustrated in Figs.~\ref{fig:MAPS_A}--\ref{fig:MAPS_G}). Namely, the precision beamsteering requirements indicated by our measurements for seamless provisioning of multi-Gbps cellular data services impose a high network signalling and control overhead. It remains to be seen whether the resulting beamsteering complexity proves prohibitive, making mm-wave deployments practically infeasible in outdoor mobile environments.

\FloatBarrier
%===========================================================================================================================
\section{Conclusions}\label{SECconclusions}
We presented results of detailed angular measurements of the signal strength of directional $60~\text{GHz}$ outdoor mm-wave links, conducted at an example pico-cellular network site in a mixed-use urban environment typical of European cities. Our results show that, although multi-Gbps mm-wave links are achievable for a substantial number of antenna orientations, mm-wave connectivity is highly site-specific and sensitive to orientation: a beam misalignment of only $10^{\circ}$ can degrade the achievable data rate by $20-100\%$. Therefore, our results reveal that the beamsteering requirements for seamless multi-Gbps mobile data provisioning are very stringent. This suggests that it would be necessary to very frequently resteer the antenna beams to compensate for slight user movements and maintain QoS; the associated network control and signalling overhead may make mm-wave deployments unattractive for outdoor mobile environments. Our ongoing work is studying the network design opportunities and limitations of mm-wave networks, via further measurements and ray-tracing simulations.

\section*{Acknowledgement}
We thank S. Katsanevakis for help with construction of the RX rotation platform and P. M\"ah\"onen for useful discussions.

\bibliographystyle{IEEEtran}
\bibliography{IEEEabrv,./mmwINFOCOMwshp2016_cameraReady}
\end{document}